\documentclass[10pt,journal,compsoc]{IEEEtran}

\usepackage[ruled,linesnumbered,lined,noend]{algorithm2e}
\usepackage[switch]{lineno}
\usepackage{cite}
\usepackage{graphicx}
\usepackage{subfigure}
\usepackage{pifont} 
\usepackage{url}
\usepackage{color}
\usepackage{booktabs}
\usepackage{multirow}
\usepackage{latexsym,bm,amsmath,amssymb}
\usepackage{array}
\usepackage{algpseudocode}
\usepackage{amsmath}
\usepackage{lscape}
\usepackage{comment}
\usepackage{balance}
\usepackage[colorlinks,linkcolor = blue,citecolor = blue,citecolor = blue,filecolor = blue,urlcolor=blue]{hyperref}
\usepackage{tcolorbox}
\usepackage{ragged2e}
\usepackage{listings}

\newcommand{\yg}[1]{\textcolor{black}{#1}}

\definecolor{light-gray}{gray}{0.95}
\newcommand{\code}[1]{\colorbox{light-gray}{\texttt{#1}}}

\newcommand{\tool}{\textsf{COTTON}}  
\newcommand{\ellLM}{{$\ell$LM}}
\newcommand{\dataset}{CodeCoT-9k}

\definecolor{light-gray}{gray}{0.85}

\definecolor{mygray}{gray}{.9}
\begin{document}
%

\title{
	Chain-of-Thought in Neural Code Generation: From and For Lightweight Language Models 
	}


\author{Guang Yang,
        Yu Zhou,
        Xiang Chen,
        Xiangyu Zhang,
        Terry Yue Zhuo,
        Taolue Chen
        
\IEEEcompsocitemizethanks{
\IEEEcompsocthanksitem Guang Yang is with the College of Computer Science and Technology, Nanjing University of Aeronautics and Astronautics, 
Nanjing, China.
E-mail: novelyg@outlook.com
\IEEEcompsocthanksitem Yu Zhou (Corresponding author) is with the College of Computer Science and Technology, 
Nanjing University of Aeronautics and Astronautics, Nanjing, China.
E-mail: zhouyu@nuaa.edu.cn
\IEEEcompsocthanksitem Xiang Chen is with the School of Information Science and Technology, Nantong University, China.
E-mail: xchencs@ntu.edu.cn
\IEEEcompsocthanksitem Xiangyu Zhang is with the College of Computer Science and Technology, 
Nanjing University of Aeronautics and Astronautics, Nanjing, China.
E-mail: zhangx1angyu@nuaa.edu.cn
\IEEEcompsocthanksitem Terry Yue Zhuo is with Monash University and CSIRO's Data61.
E-mail: terryzhuo25@gmail.com
\IEEEcompsocthanksitem Taolue Chen (Corresponding author) is with School of Computing and Mathematical Sciences, Birkbeck, University of London, UK. 
E-mail: t.chen@bbk.ac.uk
}
\thanks{Manuscript received April 19, 2020; revised August xx, xxxx.}}

\markboth{IEEE Transactions on Software Engineering,~Vol.~14, No.~8, August~2015}%
{Shell \MakeLowercase{\textit{et al.}}: Bare Demo of IEEEtran.cls for Computer Society Journals}

\IEEEtitleabstractindextext{
\begin{abstract}
\justifying
Large Language Models (LLMs) have demonstrated remarkable potential in code generation. The integration of Chain of Thought (CoT) reasoning can further boost their performance. However, current CoT methods often require manual writing or LLMs with over 100 billion parameters to generate, impeding their applicability in resource-constrained scenarios. 
In this study, we investigate lightweight Language Models  (\ellLM s), which are defined to have fewer than 10 billion parameters. 
Empirically, we find that most \ellLM s cannot generate high-quality CoTs when prompted by the few-shot method, but can take advantage of high-quality CoTs generated elsewhere to improve their performance in code generation. Based on these findings, we design a novel approach {\tool} which can leverage \ellLM s to automatically generate CoTs for code generation. 
We synthesize new datasets and conduct extensive experiments on various benchmarks. The results show that the CoTs generated by {\tool} outperform the baselines in terms of automated and human evaluation metrics.
In particular, the CoTs generated by {\tool} boost various \ellLM s to achieve higher performance gains than those generated by LLMs such as ChatGLM (130B), and are competitive with those generated by Gemini and gpt-3.5-turbo.
The results also reveal that {\tool} not only improves the performance of \ellLM s, but also enhances the performance of LLMs. 
Our study  showcases the potential of \ellLM s in software engineering applications.

\justifying
\end{abstract}

}

\maketitle

\IEEEdisplaynontitleabstractindextext
\IEEEpeerreviewmaketitle

\section{Introduction}
\label{sec:intro}


Neural code generation, which can automatically generate programs from natural language requirements based on deep learning, has become a promising approach to meet the challenges of the ever-increasing complexity of software and alleviate the burden on programmers~\cite{svyatkovskiy2020intellicode,li2022competition}. Recently, large language models (LLMs), such as GPT4~\cite{poldrack2023ai}, have demonstrated impressive performance in code generation tasks~\cite{liu2023your}.
The state-of-the-art LLMs normally have over 100 billion parameters, making even their deployment highly non-trivial. These LLMs pose challenges in terms of time, computational, and financial costs when applied to code generation, rendering them impractical for most individual users, or in resource-constrained scenarios, such as restricted access to LLM APIs or constrained GPU availability.~\cite{kaddour2023challenges, nazir2023comprehensive}. 
For software engineering applications, it is imperative to develop \emph{lightweight} language-model-based techniques that are more friendly for users (e.g., individual end users).  
Fu et al.~\cite{fu2023specializing} defined models with parameters greater than 100B as large models and those with parameters less than 10B as small models. \yg{Admittedly, the precise definition of large and small models is debatable and may evolve with the advance of technology.}
In this study, \yg{we define (pre-trained) language models (LMs) with less than 10 billion parameters as \emph{lightweight Language Models} (\ellLM), the rationale of which is that these models can be deployed on a single user graphics card (e.g., RTX 3090 or RTX 4090) based on the current technology.}
The general aim is to develop techniques 
to tackle software engineering challenges based on \ellLM s but with   
competitive performance as state-of-the-art LLMs, which would 
enable efficient, yet more accessible, software engineering applications. 

\begin{figure*}[htbp]
	\centering
	\subfigure[Evaluation on \ellLM s without chain-of-thought]{\includegraphics[width=0.48\textwidth]{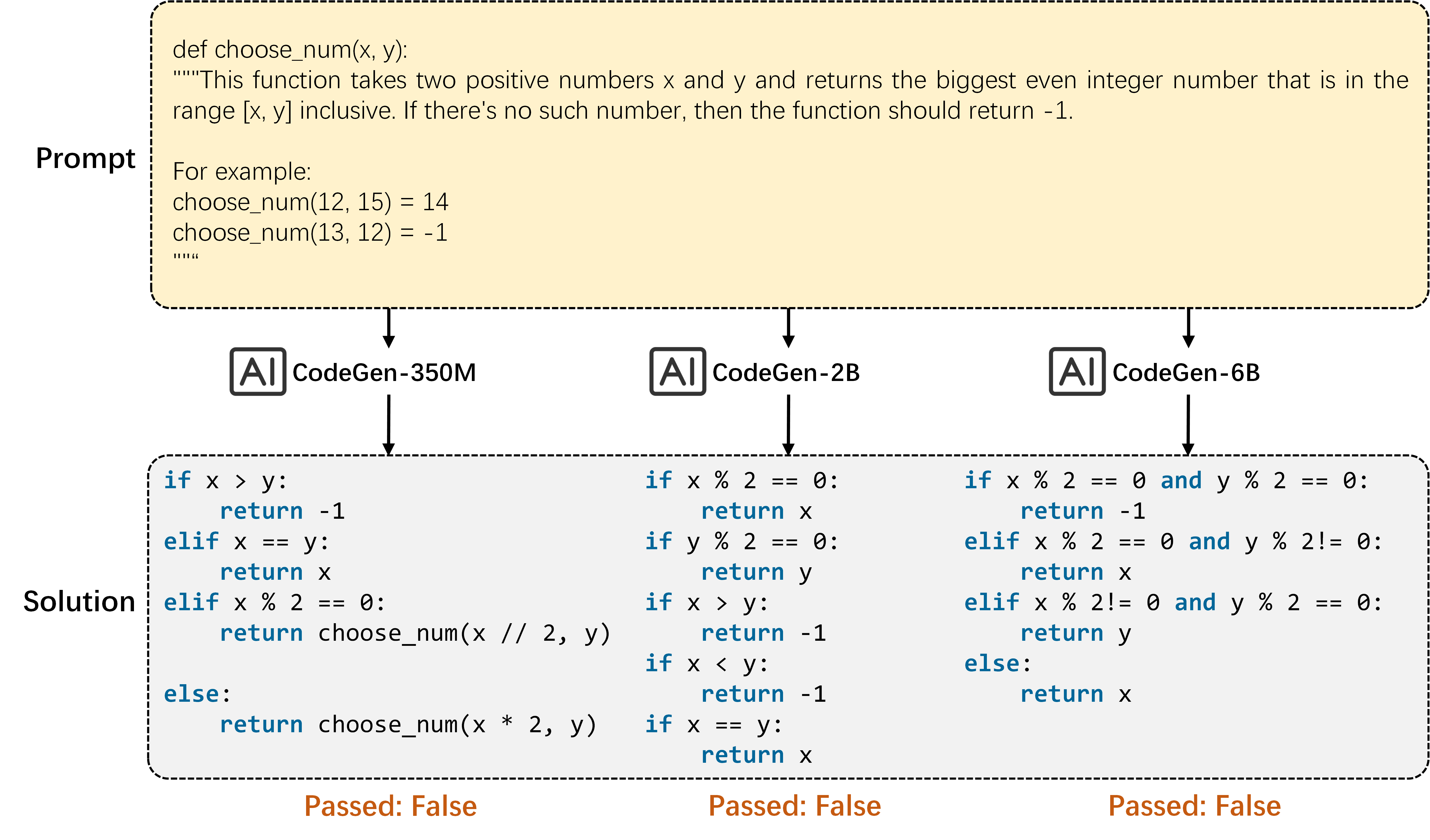}
		\label{fig:motivation-a}}
	\hfill
	\subfigure[Evaluation on \ellLM s with chain-of-thought]{\includegraphics[width=0.48\textwidth]{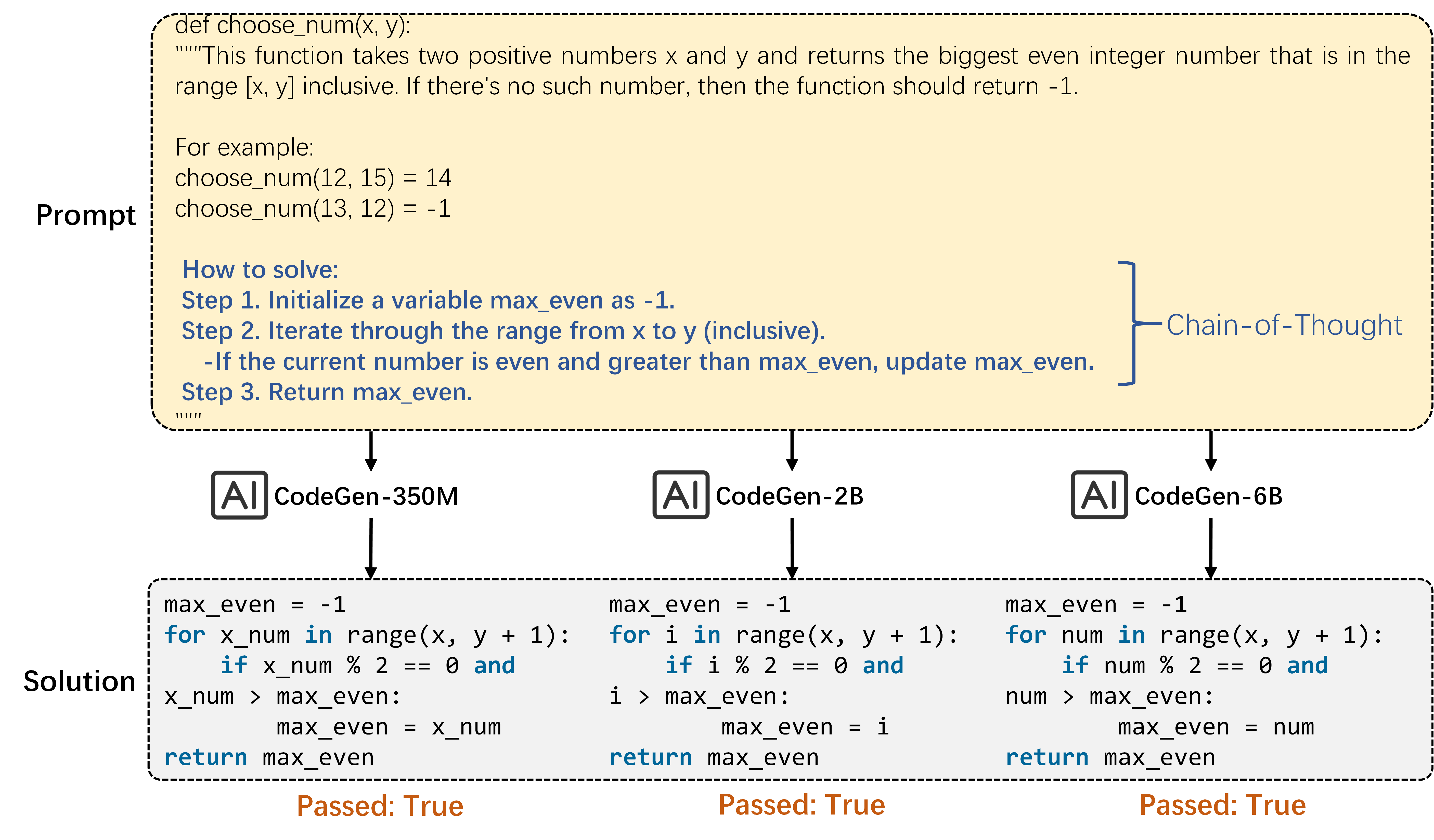}
		\label{fig:motivation-b}
	}
	\caption{The motivating examples illustrating the potential of using chain-of-thought for \ellLM s in code generation	}
 \label{fig:motivation}
\end{figure*}

Recent studies~\cite{liu2023improving, nashid2023retrieval, cao2023study, white2023chatgpt} have highlighted the importance of enhancing LLM performance by providing adequate information in the prompts. 
To improve LLMs without retraining or fine-tuning, 
researchers have resorted to Chain of Thought (CoT) techniques~\cite{wei2022chain}. 
A CoT is, in a nutshell, a series of intermediate natural language reasoning steps that lead to the final output, which enables LLMs to provide more reliable answers through thoughtful consideration and explanation.
CoT techniques have shown effectiveness in logical reasoning tasks by breaking them down into understandable intermediate steps, enabling LLMs to handle each step individually. This process not only enhances model performance but also offers the potential for model interpretability. 

Inspired by the success of CoT techniques in logical reasoning, researchers have explored their application in the code generation task.
For example, Jiang et al.~\cite{jiang2023self} proposed a self-planning approach. 
Li et al.~\cite{li2023enabling} introduced a structured CoT approach to assist models in understanding complex intentions and reducing problem-solving difficulties. 
Zhuo~\cite{zhuo2023large} introduced an evaluation metric for code generation based on LLMs and demonstrated that CoT can enhance evaluation reliability.

The previous research  has primarily focused on investigating the impact of CoT on LLMs, leaving questions regarding whether \ellLM s can also benefit from the guidance of CoT.
In \figurename~\ref{fig:motivation}, we present a motivation example to demonstrate the potential of CoT for \ellLM s in code generation.
Specifically, in \figurename\ref{fig:motivation-a}, the programming task is \code{choose\_num}, which takes two positive numbers $x$ and $y$ and returns the largest even integer that falls into the interval $[x,y]$. 
The example highlights that the original prompts for the \ellLM s (CodeGen-350M, CodeGen-2B, and CodeGen-6B) fail to generate correct code solutions.
However, by leveraging CoT in \figurename\ref{fig:motivation-b}, we modify the original prompt by using "How to solve:" and breaking down the problem into multiple steps, where the natural language explanations guide the model's understanding of the task, including instructions on branching and looping structures. With the new CoT, these \ellLM s can generate correct code solutions.

Furthermore, 
previous studies possess certain limitations as the current methods for CoT generation heavily rely on manual writing of CoTs or the utilization of LLMs~\cite{huang2022towards, qiao2022reasoning}, leading to high costs.
%
These limitations 
motivate us to investigate the following two main questions. 
(1) Can \ellLM s~independently generate high-quality CoTs to guide code generation, 
and (2) can \ellLM s 
benefit from generated CoTs?  
Here, ``independently'' 
means no model training or model parameter updating.

\smallskip
\noindent\textbf{Empirical observations.} 
To address the first question, we conduct empirical studies on the CoT generation capabilities of 11 different \ellLM s and two LLMs. 
We adopt a zero-shot approach~\cite{kojima2022large} and some few-shot approaches (such as Self-planning~\cite{jiang2023self}, SCoT~\cite{li2023enabling}, and the self-cot we propose), which provide \ellLM s with a set of examples to generate the corresponding CoT.
Our finding shows that most \ellLM s with parameter scales ranging from 0.3 to 7 billion, unfortunately, did \emph{not} demonstrate the ability to generate high-quality CoTs independently (cf.\ Section~\ref{sect:rq1} for details). 
To address the second question, we 
compare the performance of \ellLM s in code generation \emph{with} and \emph{without} CoTs.
Our findings suggest that 
all \ellLM s 
obtain performance improvement with the CoTs. 
As an example, the performance of the CodeT5 + 6B model on the HumanEval-plus dataset~\cite{liu2023your} can be improved from 26.83\% to 43.90\% with the CoTs generated by our methods (cf.~Section~\ref{sect:rq3} for details).  

\figurename~\ref{fig:motivation} provides  a motivation example, 
where  CodeGen~\cite{nijkamp2022codegen} is used as a case study. We evaluate its performance by considering varying parameter sizes 350M, 2B, and 6B. 
Without CoT, 
these models did not generate the correct code (cf.\ Fig~\ref{fig:motivation}(a)). 
However, with CoT, we decompose user requirements into three intermediate steps.
In the first step, we initialize a variable \code{max\_even} as -1;
in the second step, we define the details of loop conditions and judgment conditions;
in the third step, we return the value.
As such, we can effectively instruct the models on the necessary actions at each step, and they
eventually generate semantically correct code (though these variables have different names). 

\smallskip

\noindent\textbf{Technical contributions.} Based on the empirical observations, a natural question is how to enable \ellLM s to 
generate meaningful CoTs for code generation. 
To this end, we design a novel approach {\tool} (\textbf{C}hain \textbf{O}f \textbf{T}hough\textbf{T} c\textbf{O}de ge\textbf{N}eration).  
Specifically, {\tool} contains data collection, model training, and model inference steps.
To build the corpus, we first mine shared open source datasets (such as TheVault~\cite{manh2023vault}) to gather pairs of natural language and programming language. 
Then, we improve the dataset quality by using carefully designed heuristic cleaning rules. 
To ensure the quality of CoTs in the corpus, we use ChatGPT as the base agent and propose a multi-agent alignment method to construct high-quality CoTs (details in Section~\ref{sec:data}). 
Finally, our collected {\dataset} consists of 9,264 data pairs.


For model training, we employ CodeLlama-7b{\footnote{\url{https://github.com/facebookresearch/codellama}} as the base model to generate CoTs automatically based on the given prompt.
CodeLlama-7b incorporates advanced techniques (such as RMSNorm~\cite{zhang2019root} and Group Query Attention~\cite{ainslie2023gqa}), which enhance its performance beyond the Transformer~\cite{vaswani2017attention}. 
By applying these techniques, we can further improve the performance of {\tool}. 
To reduce the training cost, we adopt instruction-tuning and LoRA techniques~\cite{hu2021lora} to fine-tune the model parameters. This approach allows {\tool} to be trained efficiently on a single consumer graphics card while maintaining its performance.



\medskip

\noindent\textbf{Evaluation.} We conduct a comprehensive evaluation of the quality of the CoTs generated by 
{\tool} 
on the HumanEval benchmark~\cite{chen2021evaluating}.
To ensure the generalizability of {\tool}, we further collected a new code generation dataset OpenEval, and evaluated {\tool} on the OpenEval benchmark as well.
Specifically, we selected nine other commonly used models as base models and compared the results with the same training process. The quality of CoTs generated by {\tool} is superior to others in both automated and human evaluation metrics.
Furthermore, we evaluate the performance 
on \ellLM s when adopting the generated CoTs on code generation benchmarks (such as HumanEval, HumanEval-plus, and OpenEval). 
The results show that, for various \ellLM s, the CoTs generated by {\tool} achieve higher performance gains than those generated by LLMs such as ChatGLM (130B)~\cite{zeng2022glm}, and are competitive with those generated by Gemini and gpt-3.5-turbo.

Taking the CodeT5+ 6B model as an example, we observed significant performance improvements on the HumanEval, HumanEval-plus, and OpenEval benchmarks by incorporating {\tool}.
For the pass@1 metric on HumanEval and HumanEval-plus, {\tool} enhances the performance of the CodeT5+ 6B model from 26.22\% and 26.83\% to 42.68\% and 43.90\% respectively. In comparison, the LLM ChatGLM 130B only achieves an improvement of 36.59\%.
Similarly, on the OpenEval benchmark, {\tool} boosts the pass@1 metric of the CodeT5+ 6B model from 20.22\% to 35.39\%. Meanwhile, the LLM ChatGLM 130B achieves an improvement of only about 32.02\%.

In addition, to further evaluate the capabilities of {\tool}, we conducted experiments to assess its impact on LLMs (such as gpt-3.5-turbo). The results show that gpt-3.5-turbo could show a significant performance improvement in code generation when guided by the CoTs generated by {\tool} where the performance even exceeds that of the GPT-4 zero-shot scenario on the HumanEval dataset.

\yg{Finally, to demonstrate the effectiveness of {\tool} compared to fine-tuning, we utilize the StarCoder-series models as examples. The results indicate that the combination of StarCoder-7B with {\tool} has already exceeded the performance of StarCoder-16B in zero-shot scenarios and can even achieve  comparable results to a fine-tuned StarCoder-16B model.
The efficacy of {\tool} in enhancing performance across multiple models without the necessity of 
fine-tuning individual model is noteworthy, as it not only saves time and computation resources for model construction but also 
opens an inviting avenue to avoid traditional fine-tuning in adapting language models.}

In summary, the main contributions of our study can be summarized as follows:

\begin{itemize}
\item We empirically show that most existing \ellLM s lack the capability to generate high-quality CoT independently. 

\item We design a novel approach {\tool} to generate high-quality CoTs for guiding code generation, the efficacy of which has been confirmed by extensive experiments on a comprehensive set of benchmarks. 



\item We construct a new dataset of high-quality CoTs, i.e., {\dataset}, by mining existing open-source datasets. Moreover, we also construct OpenEval\footnote{\url{https://github.com/NTDXYG/open-eval}}, another dataset to benchmark the code generation performance. These datasets could be reused in similar software engineering tasks.

\end{itemize}

To facilitate the replication of {\tool}, we make our source code, trained models, and datasets publicly available on GitHub.\footnote{\url{https://github.com/NTDXYG/COTTON}}

\medskip

\noindent\textbf{Structure of the paper.} The rest of the paper is organized as follows. 
Section~\ref{sec:preliminaries} provides preliminary knowledge related to our study.
Section~\ref{sec:method} describes the framework of {\tool} and its key components. 
Section~\ref{sec:setup} and~\ref{sec:result} present the experimental design and the result analysis respectively.
Section~\ref{sec:discussion} further discusses our approach followed by the related work review in Section~\ref{sec:related}. 
Section~\ref{sec:conclusion} concludes our study and outlines possible future directions.

\section{Preliminaries}
\label{sec:preliminaries}

In this section, we first formulate the code generation task. Then we provide the fundamental concepts of our used base model CodeLlama.

\subsection{Code Generation Task Formulation}

Let $\mathcal{D} = \{(X_i,Y_i)\}_{i=1}^{|\mathcal{D}|}$ denote a code generation dataset, comprising $|\mathcal{D}|$ pairs $(X_i,Y_i)$. Here, $X_i$ represents the functional description, and $Y_i$ represents the corresponding code snippet. 
Neural code generation model $M_{code}$ aims to generate $Y_i$ conditioned on  $X_i$.
This autoregressive generation process is parameterized by $\theta_{code}$, and can be expressed as
\[  
P_{\theta_{code}}(Y_i|X_i)=\prod_{k=1}^{n}P_{\theta_{code}}
(Y_{i,k}|X_i,Y_{i,1}:Y_{i,k-1}) 
\]
where $Y_{i,1}:Y_{i,k-1}$ represents the previous sequence before the $k$-th token of $Y_i$, and $n$ denotes the number of tokens in the target sequence $Y_i$. 

To improve the code generation performance, we utilize a CoT generation model, denoted as $M_{cot}$, which generates high-quality CoT $C_i$ based on $X_i$.
Then the original input sequence $X_i$ will be augmented by concatenating with the generated CoT $C_i$, resulting in a new input sequence $\hat{X}_i = X_i \oplus C_i$, where $\oplus$ denotes the concatenation operation. 
Subsequently, we approximate the probability of generating the code snippet $Y_i$ given the input sequence $X_i$ as
\[  
  P(Y_i|X_i) \propto \underbrace{P_{\theta_{cot}}(C_i|X_i)}_{M_{cot}} \underbrace{P_{\theta_{code}}(Y_i|X_i, C_i)}_{M_{code}}
\] 
%
In our study, we treat the existing neural code generation model $M_{code}$ as a black box. 
We intend to train a model $M_{cot}$ to generate CoTs, which can be used to guide code generation and further improve the performance of this task. 

\subsection{Code Language Model CodeLlama}

CodeLlama~\cite{roziere2023code} is a code language model built on Llama-2~\cite{touvron2023llama}, which stands out for its exceptional performance in open models, padding capabilities, support for large input contexts, and zero-sample instruction tracking for programming tasks.
In our study, we use CodeLlama-7B as the base model for {\tool} due to its remarkable code understanding and generation capabilities. 

\smallskip 

\noindent\textbf{Embedding Layer.} 
CodeLlama tokenizes the given functional description $X$ into sub token sequences $\left\{ w_{i}\right\} _{i=1}^{N}$ using byte-pair encoding (BPE) and the SentencePiece algorithm~\cite{yang2022survey}.
Each sub-word $w_i$ is then transformed into an embedding (row) vector $\boldsymbol{x}_{i} \in \mathbb{R}^{d}$, where $d$ represents the dimension of the embedding vector.
These embedding vectors are combined into a matrix $\boldsymbol{X} = \left\{ \boldsymbol{x}_{i}\right\} _{i=1}^{N}$, which represents the meaningful relationship between the tokens in the input sequence.
By using this embedding matrix, {\tool} captures the semantic information of the functional description and prepares it for further processing in the subsequent layers of the model.

\noindent\textbf{RMSNorm.} 
CodeLlama utilizes Root Mean Square Layer Normalization (RMSNorm) instead of LayerNorm for normalization purposes~\cite{zhang2019root}.
RMSNorm operates by normalizing each embedding vector $\boldsymbol{x}_i$ by dividing it through the root mean square. This normalization process helps reduce the impact of noise and improves computational efficiency.
For each embedding vector $\boldsymbol{x}_{i}$, the calculation formula of RMSNorm is defined as follows.
\begin{displaymath}
\boldsymbol{x}_{i}=\dfrac{\boldsymbol{x}_{i}}{RMS\left( \boldsymbol{X}\right) }\cdot g_{i}
\end{displaymath}
where $RMS\left( \boldsymbol{X}\right) = \sqrt{\dfrac{1}{n}\sum ^{n}_{i=1}\boldsymbol{x}_{i}^{2}}$ represents the root mean square of the embedding matrix $\boldsymbol{X}$, and $g_{i}$ denotes the rescale factor.

\noindent\textbf{Group Query Attention (GQA).} 
CodeLlama introduces GQA~\cite{ainslie2023gqa} as a modification to the standard multi-head attention mechanism. This modification optimizes the model's performance by dividing the Query heads into groups, with each group sharing the same Key and Value matrix. Moreover, the model incorporates Rotary Position Embedding (RoPE)~\cite{su2021roformer} and FlashAttention~\cite{dao2022flashattention} for further improvement.

Specifically, for the given matrix $\boldsymbol{X}$, the model computes the Query, Key, and Value matrix as follows.
\begin{displaymath}
\boldsymbol{q}_{i}=f_{q}\left( \boldsymbol{x}_{i},i\right)
\end{displaymath}
\begin{displaymath}
\boldsymbol{k}_{j}=group\left( f_{k}\left( \boldsymbol{x}_{j},j\right)\right)
\end{displaymath}
\begin{displaymath}
\boldsymbol{v}_{j}=group\left( f_{v}\left( \boldsymbol{x}_{j},j\right)\right)
\end{displaymath}
where $\boldsymbol{q}_{i}$ represents the Query vector of the embedding vector $\boldsymbol{x}_{i}$, incorporating the position information. $\boldsymbol{k}_{j}$ and $\boldsymbol{v}_{j}$ denote the Key and Value vectors of the embedding vector $\boldsymbol{x}_{j}$, respectively, incorporating the position information $j$. The grouping operation is applied to ensure that the Query heads within each group share the same Key and Value matrix.
To compute the self-attention output corresponding to the $i$-th embedding vector $\boldsymbol{x}_{i}$, an attention score is computed between $\boldsymbol{q}_{i}$ and the other $\boldsymbol{k}_{j}$ vectors. This attention score is then multiplied by the corresponding $\boldsymbol{v}_{j}$ vectors and summed to obtain the output vector. 
\begin{displaymath}
a_{i,j}=\dfrac{\exp \left( \dfrac{\boldsymbol{q}^{T}_{i}{\boldsymbol{k}_i}}{\sqrt{d}}\right) }{\sum ^{N}_{m=1}\exp \left( \dfrac{\boldsymbol{q}^{T}_{i}{\boldsymbol{k}_m}}{\sqrt{d}}\right) }
\qquad 
\boldsymbol{o}_{i}=\sum ^{N}_{j=1}a_{i,j}\cdot \boldsymbol{v}_{j}
\end{displaymath}
where $a_{i,j}$ represents the attention score and $\boldsymbol{o}_{i}$ denotes the output vector for the $i$-th embedding vector.


\noindent\textbf{FFN.} 
The Feed Forward Network (FFN) in CodeLlama consists of linear layers and an activation function. It operates on the matrix $\boldsymbol{X}$ to calculate the output using a specific formula.
\begin{displaymath}
\textbf{FFN}(\boldsymbol{X}) = f_{down}\left( f_{up}\left( \boldsymbol{X} \right) \times SiLU(f_{gate}\left( \boldsymbol{X} \right)) \right)
\end{displaymath}
where $SiLU$ represents the activation function, defined as the element-wise product of the Sigmoid function and the input. This activation function introduces non-linearity to the network.

During the overall process, CodeLlama generates the output probability $P$ for a given input $\boldsymbol{X}$ through a series of operations, including Group Query Attention, RMSNorm, and FFN.
Initially, the input $\boldsymbol{X}$ is normalized using RMSNorm. Then, GQA is applied to the normalized input and added to the original input, resulting in $\boldsymbol{X}_{hidden}$.
Next, $\boldsymbol{X}_{hidden}$ is again normalized using RMSNorm, and FFN is applied to it, obtaining $\boldsymbol{X}_{final}$.
Finally, $\boldsymbol{X}_{final}$ is normalized using RMSNorm and passed through the function $f_{vocab}$ to map it to the output probability space, resulting in the final output probability $P$.
\begin{displaymath}
P = f_{vocab} (\textbf{RMSNorm}(\boldsymbol{X}_{final})))
\end{displaymath}

\section{Our Approach}
\label{sec:method}

The workflow of our proposed {\tool} is shown in \figurename~\ref{fig:Workflow}.
There are three major steps, i.e., data collection, model training, and model inference. In the rest of this section, we show the details of these three steps.



\begin{figure}[htbp]
\centering
\vspace{-2mm}
\includegraphics[width=0.45\textwidth]{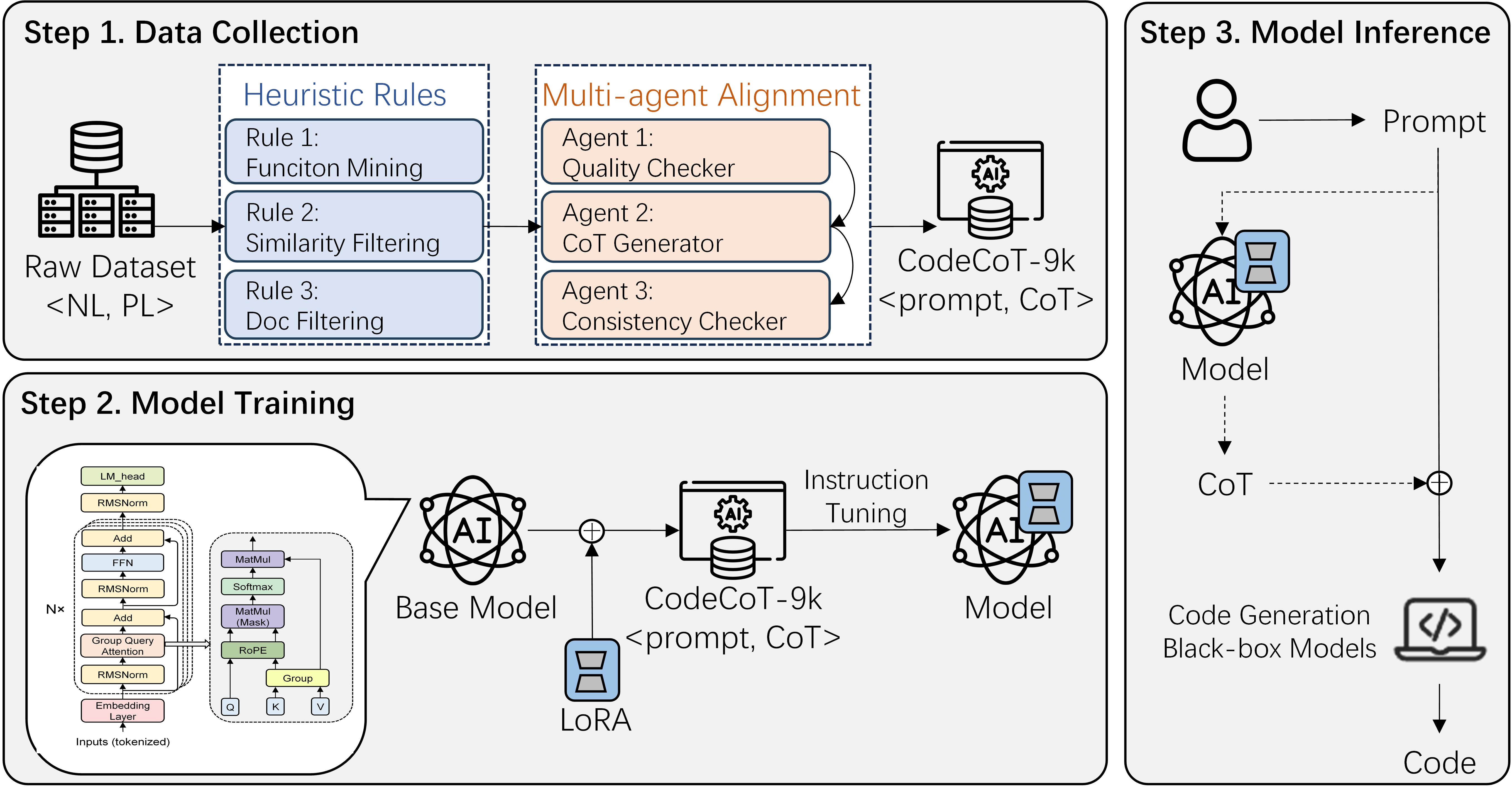}
\caption{The workflow of the proposed approach {\tool}}
\vspace{-2mm}
\label{fig:Workflow}
\end{figure}

\subsection{Data Collection}
\label{sec:data}

The construction of the CoT dataset {\dataset} follows the process shown in Step 1 in \figurename~\ref{fig:Workflow}. 
We begin by selecting TheVault\footnote{\url{https://github.com/FSoft-AI4Code/TheVault}}, MBPP\footnote{\url{https://huggingface.co/datasets/mbpp}} and LeetCode\footnote{\url{https://huggingface.co/datasets/mhhmm/leetcode-solutions-python}} as our raw datasets. 
These datasets consist of natural language descriptions of functional requirements paired with corresponding implementation code snippets. They have been widely used in the literature~\cite{hou2023large}.

However, after our manual analysis, we find that a significant portion of the code snippets in these datasets are not self-contained, requiring external modules or files for program comprehension~\cite{gunasekar2023textbooks}, which poses challenges for CoT generation.
In addition, we notice that some code snippets consist of trivial or template code (for defining constants, setting parameters, configuring GUI elements, etc), which is not useful for CoT generation.  
As a result, to improve the quality of our dataset, we design two data cleaning methods, i.e.,  heuristic rule-based cleaning and multi-agent alignment-based cleaning.

\smallskip
\noindent\textbf{Heuristic rule-based cleaning.} 
These rules are designed to filter data that contains syntactically incorrect code and documents that are inconsistent with the code, in addition to avoiding data leakage problem.
We define three heuristic rules as below. 
\begin{itemize}
\item[\textbf{R1}] Code Filtering. We utilize the AST parser tool to extract method-level code and corresponding functional comments, by which syntactically incorrect code can be filtered. 
\item[\textbf{R2}] Doc Filtering. 
We utilize DocChecker~\cite{DocChecker} to determine the consistency between the documentation and the code, which can maintain an accurate alignment between the comments and the code effectively. We then remove code snippets with inconsistent documentation.
\item[\textbf{R3}] Similarity Filtering. To prevent data leakage in the training set, We utilize the codet5p embedding model\footnote{\url{https://huggingface.co/Salesforce/codet5p-110m-embedding}} for code representation learning. We then remove the code snippets that exceed the semantic similarity threshold by considering the cosine similarity. 
\end{itemize}

\smallskip
\noindent\textbf{Multi-agent alignment-based cleaning.} 
These agents are designed to filter low-quality data, including snippets that are not educationally meaningful and CoTs that are inconsistent with the semantics of the code.
We leverage the power of multiagents 
to align and clean the data, where 
they are based on gpt-3.5-turbo.\footnote{\url{https://platform.openai.com/docs/models/gpt-3-5}} 
Specifically, we implement the agent definition by constructing the specific prompt~\cite{wang2023survey}: 

\begin{itemize}
\item[\textbf{A1}] \textbf{Quality Checker}. 
This agent assesses the educational value of the data and removes low-quality items, ensuring that the dataset comprises high-quality code snippets. 
%
\item[\textbf{A2}] \textbf{CoT Generator.} We first transform code snippets and functional comments into a standardized prompt and solution format similar to HumanEval~\cite{chen2021evaluating}. 
This agent employs a one-shot approach, providing an example to aid the agent in learning the desired output style and generating CoT based on user input.
Importantly, we intentionally do not disclose the specific implementation details of the code to the agent, which encourages the agent to generate diverse CoTs, as code implementation can vary widely. 

\item[\textbf{A3}] \textbf{Consistency Checker}. This agent examines the consistency between the CoT instructions generated by Agent 2 and the code snippets. It removes code snippets with inconsistencies, ensuring that the CoT instructions accurately reflect the code's behavior.
\end{itemize}


\begin{tcolorbox}[width=1.0\linewidth, title={Quality Checker}]
Give you a code snippet, determine its educational value for a student whose goal is to learn basic coding concepts. \\
If it has educational value, return only ``Yes", else return "No".
\end{tcolorbox}

\begin{tcolorbox}[width=1.0\linewidth, title={Consistency Checker}]
Given a piece of code and a chain of thought, determine whether they express exactly the same functional semantics.\\
If consistent, return only "Yes", else return "No".
\end{tcolorbox}

\begin{tcolorbox}[width=1.0\linewidth, title={CoT Generator}]
\#\#\# Given a piece of code, output the corresponding implementation idea.\\
\#\#\# Example:\\
Input:\\
from typing import List\\
def below\_zero(operations: List[int]) -\textgreater bool:\\
    """ You're given a list of deposit and withdrawal operations on a bank account that starts with zero balance. 
    Your task is to detect if at any point the balance of account falls below zero, and at that point function should return True. 
    Otherwise it should return False.\\
    """\\
Output:\\
How to solve:\\
Step 1. Initialize account balance as 0.\\
Step 2. Iterate through operations.\\
    -add value to account balance.\\
    -If account balance \textless 0, return True.\\
Step 3. Return False.\\
\#\#\# Input: $[X]$\\
\#\#\# Output: $[Y]$
\end{tcolorbox}

\subsection{Model Training}

To enable the training of \ellLM s with limited resources, researchers have explored parameter-efficient fine-tuning methods since full parameter fine-tuning is impractical in this scenario. 
As demonstrated in the previous study~\cite{zhuo2024astraios}, these methods have been proven to achieve high performance by providing sufficient instructions to the language models.
In contrast to continuous-based soft prompt methods~\cite{lester2021power, gu2022ppt}
, our approach employs a set of discrete tokens as instruction prompts, which are meaningful and easily interpretable. 
For our CoT generation task, we design a template that incorporates task-specific instructions, which is illustrated as follows.

\begin{tcolorbox}[width=1.0\linewidth, title={Instruction Template}]
\#\#\# Given a piece of code, output the corresponding implementation idea.\\
\#\#\# Input: $[X]$\\
\#\#\# Output: $[Y]$
\end{tcolorbox}

To address the challenge of excessive parameters in the base model, we employ the LoRA method~\cite{hu2021lora} to facilitate efficient fine-tuning with limited resources.
Unlike traditional fine-tuning methods that update all weights of the model, LoRA introduces trainable low-rank matrices to approximate weight adjustments. This approach leverages the observation that the adaptation process inherently exhibits a low ``intrinsic rank.''
Let $\boldsymbol{W_{0}} \in \mathbb{R}^{d \times k}$ denote the pre-trained matrix. The weight adjustment approximation from $\boldsymbol{W_{0}}$ to $\boldsymbol{W_{0}} + \Delta \boldsymbol{W}$ using LoRA can be expressed as:
\begin{displaymath}
\boldsymbol{W_{0}} + \Delta \boldsymbol{W} = \boldsymbol{W_{0}} + \boldsymbol{B}\boldsymbol{A}
\end{displaymath}

Here, $\boldsymbol{B} \in \mathbb{R}^{d \times r}$ and $\boldsymbol{A} \in \mathbb{R}^{r \times k}$, where $r \ll \min (d, k)$ represents the rank. 
During the fine-tuning process, $\boldsymbol{W_{0}}$ remains unchanged, while $\boldsymbol{B}$ and $\boldsymbol{A}$ become the trainable parameters. 
Given an input $\boldsymbol{X}$ and its associated original output $\boldsymbol{H}$, the adjusted output $\bar{\boldsymbol{H}}$ is computed as:
\begin{displaymath}
\bar{\boldsymbol{H}} = \boldsymbol{W_{0}} \boldsymbol{X} + \Delta \boldsymbol{W} \boldsymbol{X} = \boldsymbol{H} + \boldsymbol{B}\boldsymbol{A} \boldsymbol{X}
\end{displaymath}

To initialize the matrices, Matrix $\boldsymbol{A}$ is initialized by random Gaussian values, while $\boldsymbol{B}$ is initialized by zeros. 
This ensures that the initial value of $\Delta \boldsymbol{W} = \boldsymbol{B}\boldsymbol{A}$ is zero at the start of training.
To increase the number of trainable parameters and improve the capabilities, we apply LoRA to adapt all linear layers simultaneously.

\subsection{Model Inference} 



In the inference phase, the model trained by {\tool} can be efficiently deployed on a single consumer graphics card to generate CoT. 
Note that {\tool} is a stand-alone tool and its deployment does not require the use of LLM.
To accelerate the decoding process, {\tool} utilizes the Greedy Search algorithm, which, in a nutshell, selects the token with the highest probability at each decoding step, resulting in a more deterministic output during inference.

The generated CoT by {\tool} serves as an additional piece of information that is added to the original prompt in code generation tasks. 
In practical scenarios, to improve the user experience, we recommend generating CoT only when the code generation model fails to generate the correct code. This approach ensures that the CoT is used when necessary to avoid unnecessary overhead. 
\section{EXPERIMENTAL SETUP}
\label{sec:setup}

To evaluate the effectiveness and benefits of our proposed approach,
we mainly design the following three research questions (RQs):


\noindent\textbf{RQ1: Can \ellLM s generate high-quality CoT independently?}

In this RQ, we want to investigate whether \ellLM s have the capability to generate high-quality CoTs independently (i.e., the first question mentioned in Section~\ref{sec:intro}). A negative finding of this RQ can constitute the motivation for designing our approach \tool. 

\noindent \textbf{RQ2: Can {\tool} 
	generate higher-quality CoTs?}

%

In this RQ, we want to evaluate the effectiveness of our approach \tool. 
Specifically, we aim to compare its performance against state-of-the-art base models. 
Since we are the first to study automated CoT generation for code generation, we select relevant base models from similar research topics. 
We employ automatic evaluation metrics to assess the quality of the generated CoTs from various perspectives. 
We also conduct a human evaluation to assess the effectiveness of our approach, since automatic metrics may not fully capture the semantic similarity and educational value of the generated CoTs.

\noindent\textbf{RQ3: Can \ellLM s effectively benefit from CoT?}


While \ellLM s may not be able to 
generate high-quality CoTs independently, they may benefit from the provided CoTs to improve code generation performance (i.e., the second question mentioned in Section~\ref{sec:intro}. ). 
In this RQ, we want to validate the effectiveness of leveraging CoT for \ellLM s. 



\subsection{Dataset}

\subsubsection{Code Generation}

To evaluate the performance of \ellLM s with and without CoT in the zero-shot scenario, we conduct experiments on three 
code generation datasets. 

\noindent \textbf{HumanEval/HumanEval-plus.} 
The HumanEval dataset~\cite{chen2021evaluating} was developed and published by OpenAI, consisting of 164 Python programming problems.
Each problem includes an average of 7.8 test cases. which can provide a comprehensive evaluation of code generation capabilities.
%
The HumanEval-plus dataset~\cite{liu2023your}
aims to alleviate the test case coverage limitation in HumanEval, which may result in false positive rates. Note that HumanEval and HumanEval-plus are only different in terms of test cases and not in terms of CoT generation tasks. 

\noindent \textbf{OpenEval.} 
To ensure fairness and generalizability, we collect a new code generation dataset OpenEval. 
This dataset includes 178 problems selected from the competition-level code translation dataset AVATAR~\cite{ahmad2021avatar}. 
For each problem, we designed additional test cases in a manual way that can effectively evaluate the quality of the generated code and minimize bias and leakage.
Particularly, we hired two software engineers with 2$\sim$3 years of development experience each to construct 5 test cases for each code segment to ensure the diversity of test cases.

\subsubsection{CoT Generation}

Following the method in Section~\ref{sec:data}, we collect a total of 9,264 CoT-generated samples. 
These samples are randomly split into a training set of 9,000 samples and a validation set of 264 samples.
To evaluate the performance of {\tool}, we generate CoTs on the HumanEval and OpenEval datasets, using the same methodology described in Section~\ref{sec:data}. The derived datasets are HumanEval-CoT and OpenEval-CoT respectively.
Furthermore, we utilize Agent 2 (cf. Section~\ref{sec:data}) as the Teacher Model. 
An example in our used datasets is shown in \figurename~\ref{fig:dataset}. 
Finally, Table~\ref{tab:statistics} provides statistical information about the datasets used for our evaluation.

\begin{figure}[htbp]
\centering
\includegraphics[width=0.45\textwidth]{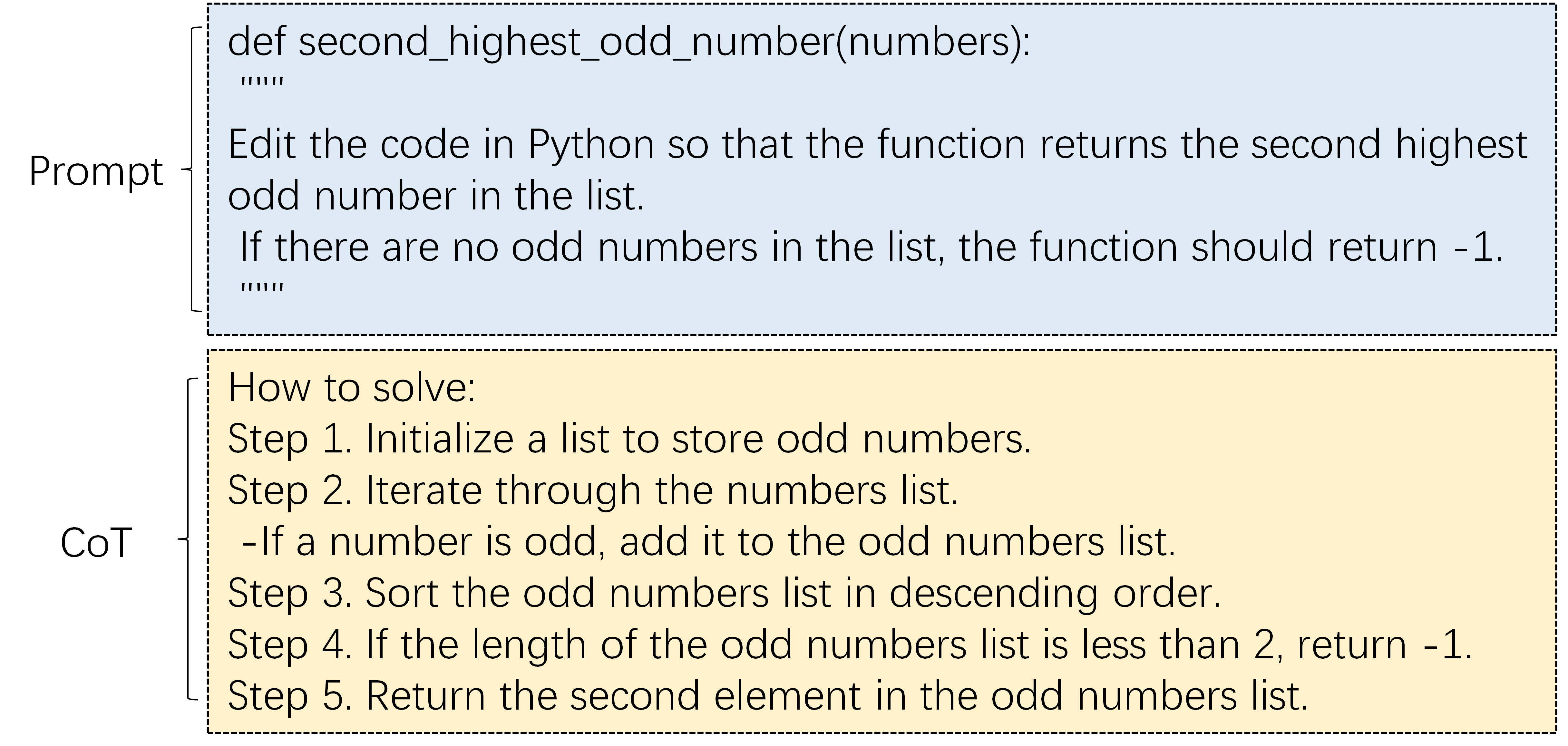}
\caption{An example in our used datasets}
\label{fig:dataset}
\end{figure}

\begin{table}[htbp]
 \caption{Statistical information of our datasets}
 \begin{center}
 \scalebox{0.85}{
\begin{tabular}{clccc}
\toprule
\textbf{Type} & \textbf{Train} & \textbf{Valid} & \textbf{HumanEval-CoT} & \textbf{OpenEval-CoT} \\ 
\midrule
Count & 9,000  & 264 & 164 & 178 \\
\midrule
Avg in Prompt & 63.76 & 61.32 & 80.09 & 79.39 \\
Median in Prompt & 41.00 & 39.50 & 64.50 & 64.50 \\
$\leq 256$ in Prompt & 98.03\% & 98.11\% & 98.78\% & 99.44\% \\
\midrule
Avg in CoT & 85.40 & 82.41 & 94.54 & 93.99 \\
Median in CoT & 74.00 & 74.50 & 86.00 & 85.00 \\
$\leq 256$ in CoT & 99.07\% & 99.24\% & 99.39\% & 99.44\% \\
  \bottomrule
\end{tabular}
}
 \label{tab:statistics}
 \end{center}
\end{table}

\subsection{Code Generation Models} \label{sect:cgm}
Based on the performance evaluations conducted by Gunasekar et al.~\cite{gunasekar2023textbooks}, we select the following state-of-the-art \ellLM s  for our experiments: CodeGen~\cite{nijkamp2022codegen}, StarCoder~\cite{li2023starcoder}, and CodeT5+~\cite{wang2023codet5+}.
These models have demonstrated promising performance in various code generation tasks~\cite{zhang2023unifying}. 

\noindent \textbf{CodeGen.}
CodeGen~\cite{nijkamp2022codegen} is an open-source large language model specifically designed for code generation, with a particular focus on multi-turn program synthesis. It enhances program synthesis by breaking down complex user intents into multiple steps, facilitating the model's understanding.
For our experiments, we select three models of different parameter sizes: 350M, 2B, and 6B.

\noindent \textbf{StarCoder.}
StarCoder~\cite{li2023starcoder} is a language model trained on a diverse corpus of source code and natural language text. Its training data covers over 80 programming languages and includes text extracted from GitHub issues, commits, and notebooks. 
We select three models of different parameter sizes: 1B, 3B, and 7B.

\noindent \textbf{CodeT5+.}
CodeT5+~\cite{wang2023codet5+} is a code large language model with an encoder-decoder architecture, capable of performing various code understanding and generation tasks. It offers flexibility by supporting different modes of operation. 
In our experiments, we select four models of different parameter sizes: 220M, 770M, 2B, and 6B.

\subsection{Evaluation Metrics}

\subsubsection{Code Generation}

To evaluate the performances of code generation models, we employ the Pass@1 metric and CoT-Pass@1 metric.

\noindent \textbf{Pass@1.} 
The Pass@1 metric measures the percentage of generated code snippets that pass the corresponding test cases without considering the CoT. This metric evaluates the ability of the code generation model to generate functionally correct code.

\noindent \textbf{CoT-Pass@1.} 
When the code generated by the model without the guidance of the CoT fails to pass the corresponding test cases, the model is provided with the CoT guidance to generate code. 
The CoT-Pass@1 metric is used to measure the percentage of generated code snippets that successfully pass the test cases after considering the CoT. 
This metric specifically evaluates the model's capability to generate code that passes the test cases when guided by the CoT. 
It focuses on assessing the model's ability to improve its code generation performance by incorporating the guidance provided by the CoT in cases where the initial code generation without CoT fails to pass the test cases.

\subsubsection{CoT Generation}
To evaluate the performance of the CoTs generated by {\tool}, we first employ four automatic evaluation metrics commonly used in similar generation tasks (such as code generation~\cite{yang2023exploitgen, yang2023syntax}, code summarization~\cite{yang2022ccgir, yang2022dualsc}, and code translation~\cite{yang2023assessing} tasks). 

\noindent \textbf{BLEU} (Bilingual Evaluation Understudy)~\cite{papineni2002bleu} is a machine translation metric that measures the lexical similarity between two texts by computing the overlap of $n$-grams. In our evaluation, we utilize BLEU-1, -2, -3, -4 to evaluate the quality of the generated CoTs.

\noindent \textbf{METEOR} (Metric for Evaluation of Translation with Explicit ORdering)~\cite{banerjee2005meteor} is an enhanced automatic evaluation metric based on BLEU. It incorporates an alignment algorithm based on dictionaries and linguistic knowledge, placing greater emphasis on word order and grammatical structure matching.

\noindent \textbf{ROUGE-L} (Recall-Oriented Understudy for Gisting Evaluation)-L~\cite{lin2004rouge}  measures 
the similarity between the generated CoT and the ground-truth CoT by comparing their longest common subsequence. This metric is advantageous for handling long sequences and is not limited to single vocabulary items.

\noindent \textbf{Consistency} To further automatically assess the semantic correctness of the generated CoT, we use Agent 3 (cf.\ Section~\ref{sec:data}) to examine the consistency between the CoT and the code snippets.


The values of these performance metrics range from 0 to 1 and are displayed as percentages.  
A higher value indicates a closer match between the generated CoT and the ground-truth CoT.
To compute BLEU, METEOR, and ROUGE-L, we utilize the nlg-eval library.\footnote{\url{https://github.com/Maluuba/nlg-eval}}

\subsection{Implementation Details and Running Platform}

The hyper-parameters are tuned according to actual performance. We show the values of these hyper-parameters in Table~\ref{Hyper-parameters}. 
For the implementation of {\tool} and other base models, we utilize the PyTorch\footnote{\url{https://pytorch.org/}} and Transformers\footnote{\url{https://github.com/huggingface/transformers}} libraries.
\begin{table}[ht]
    \centering
    \caption{Hyper-parameters and their values}
    \begin{tabular}{c|c||c|c}
    \toprule
       Hyper-parameter  & Value &  Hyper-parameter  & Value\\
     \midrule
      Optimizer & AdamW & Random Seed & 42 \\
      Learning Rate  & 1e-4 & Training batch size & 1 \\
      Lora R & 8 & Lora alpha & 16 \\
      Max input length & 256 & Max output length & 256 \\
      Epoch & 20 & Early Stop & 5 \\
      \bottomrule
    \end{tabular}
    \label{Hyper-parameters}
\end{table}
Our implementation is based on PyTorch 1.8, and the experiments are conducted on a machine with an Intel(R) Xeon(R) Silver 4210 CPU and the GeForce RTX 3090 GPU. It took about 6 hours to complete the model training of {\tool}.

\section{Experimental Result Analysis}
\label{sec:result}

\begin{table*}[ht]
 \caption{Performance comparison of \ellLM s and LLMs to independently generate CoTs on the HumanEval-CoT and OpenEval-CoT datasets.}
     \centering
\scalebox{0.92}{
\begin{tabular}{c|c|c|c|ccccccc}
  \toprule
\textbf{Corpus} & \textbf{Type} & \textbf{Model} & \textbf{Model Size} & \textbf{BLEU-1} & \textbf{BLEU-2} & \textbf{BLEU-3} & \textbf{BLEU-4} & \textbf{Meteor} & \textbf{Rouge-L} & \textbf{Consistency}\\
  \midrule
\multirow{13}{*}{HumanEval-CoT} 
& \multirow{12}{*}{\ellLM} 
& CodeGen & 350M & 16.39 & 11.68 & 8.79 & 6.86 & 13.08 & 23.56 & 0.61\\
& & CodeGen & 2B & 40.34 & 30.90 & 24.16 & 19.25 & 23.60 & 37.95 & 39.02 \\
& & CodeGen & 6B & 27.48 & 20.33 & 15.54 & 12.19 & 18.95 & 31.37 & 28.66 \\
& & StarCoder & 1B & 38.80 & 29.79 & 23.18 & 18.33 & 24.46 & 38.83 & 31.71 \\
& & StarCoder & 3B & 42.13 & 32.16 & 25.04 & 19.97 & 25.11 & 40.53 & 60.96 \\
& & StarCoder & 7B & 43.53 & 34.20 & 27.16 & 22.02 & 26.58 & 43.76 & 78.66 \\
& & CodeT5+ & 220M & 9.37 & 5.88 & 4.18 & 3.13 & 5.79 & 8.33 & 0.61 \\
& & CodeT5+ & 770M & 22.25 & 16.28 & 12.33 & 9.51 & 17.76 & 24.73 & 0.61\\
& & CodeT5+ & 2B & 35.77 & 27.89 & 22.09 & 17.77 & 22.55 & 35.17 & 27.44\\
& & CodeT5+ & 6B & 19.37 & 13.18 & 9.56 & 7.28 & 14.05 & 22.99 & 7.32\\
& & CodeLlama & 7B & 44.88 & 36.13 & 29.10 & 23.80 & 27.34 & 43.49 & 82.32\\
\cline{2-11}
& \multirow{5}{*}{LLM} 
& InternLM & 123B & 52.08 & 42.23 & 34.50 & 28.67 & 32.65 & 43.09 & 89.63 \\
& & ChatGLM & 130B & 53.28 & 43.28 & 35.52 & 29.64 & 32.61 & 43.03 & 86.59 \\
& & Gemini & Not-Available & 52.50 & 43.56 & 36.43 & 30.92 & 32.14 & 53.28 & 91.46\\
& & gpt-4 & Not-Available & \textbf{60.14} & \textbf{50.49} & \textbf{43.03} & \textbf{37.24} & \textbf{34.55} & \textbf{54.32} & \textbf{96.95}\\
& & Teacher(gpt-3.5-turbo) & Not-Available & - & - & - & - & - & - & 93.29\\
  \midrule
\multirow{13}{*}{OpenEval-CoT}
& \multirow{12}{*}{\ellLM} 
& CodeGen & 350M & 31.74 & 25.32 & 21.11 & 17.96 & 17.62 & 29.29 & 0.56\\
& & CodeGen & 2B & 39.22 & 31.21 & 25.22 & 20.73 & 24.27 & 37.43 & 29.78\\
& & CodeGen & 6B & 32.54 & 25.84 & 20.92 & 17.21 & 20.79 & 33.15 & 27.53\\
& & StarCoder & 1B & 41.33 & 32.09 & 25.32 & 20.44 & 24.67 & 35.98 & 21.35\\
& & StarCoder & 3B & 44.85 & 35.90 & 29.17 & 24.21 & 26.45 & 38.74 & 43.26\\
& & StarCoder & 7B & 46.23 & 36.98 & 30.07 & 24.88 & 28.35 & 43.33 & 57.87\\
& & CodeT5+ & 220M & 14.70 & 9.98 & 7.88 & 6.59 & 9.13 & 10.56 & 0.00\\
& & CodeT5+ & 770M & 29.35 & 22.95 & 18.11 & 14.52 & 19.96 & 28.59 & 0.00\\
& & CodeT5+ & 2B & 29.96 & 24.07 & 19.61 & 16.27 & 21.31 & 32.27 & 21.91\\
& & CodeT5+ & 6B & 26.58 & 21.00 & 17.20 & 14.42 & 17.13 & 28.12 & 5.06\\
& & CodeLlama & 7B & 52.09 & 43.89 & 37.27 & 32.13 & 30.52 & 48.96 & 71.91\\
\cline{2-11}
& \multirow{5}{*}{LLM} 
& InternLM & 123B & 55.59 & 46.77 & 40.01 & 34.80 & 35.28 & 46.48 & 84.27\\
& & ChatGLM & 130B & 55.28 & 46.47 & 39.68 & 34.56 & 35.14 & 45.64 & 76.40\\
& & Gemini & Not-Available & 57.49 & 49.04 & 42.43 & 37.25 & 34.13 & 54.88 & 81.46\\
& & gpt-4 & Not-Available & \textbf{62.54} & \textbf{54.34} & \textbf{47.88} & \textbf{42.81} & \textbf{36.84} & \textbf{58.02} & 87.64\\
& & Teacher(gpt-3.5-turbo) & Not-Available & - & - & - & - & - & - & \textbf{89.33}\\
  \bottomrule
\end{tabular}
}
 \label{tab:RQ1}
\end{table*}

\subsection{RQ1: Can \ellLM s generate high-quality CoTs independently?} \label{sect:rq1}

To investigate whether \ellLM s can independently generate high-quality CoTs for code generation, we conduct experiments 
based on few-shot prompt learning techniques as described in Section~\ref{sec:data}. 
We evaluate the performance of \ellLM s with model sizes ranging from 0.35B to 7B, the base model CodeLlama, and 
several representative LLMs (such as InternLM 123B~\cite{2023internlm}, ChatGLM 130B~\cite{zeng2022glm}, Gemini~\cite{team2023gemini}, gpt-3.5-turbo, and gpt-4~\cite{achiam2023gpt}). Among these LLMs, Gemini, gpt-3.5-turbo, and gpt-4 are currently considered to be the most promising LLMs in code generation tasks and have shown promising performance.

Table~\ref{tab:RQ1} presents the performance of \ellLM s and LLMs across all evaluation metrics on the HumanEval-CoT and OpenEval-CoT datasets.
In terms of lexical similarity, \ellLM s generally performed worse than LLMs. For example, using the \textbf{METEOR} metric, LLMs like gpt-4 can achieve scores around 0.35 on the HumanEval-CoT dataset and around 0.37 on the OpenEval-CoT dataset. In contrast, the majority of \ellLM s scored below 0.3 on both datasets.
In terms of semantic perspective, most \ellLM s are difficult to generate high-quality CoTs, while LLMs demonstrated better performance. For example, based on the \textbf{Consistency} metric, gpt-4 can achieve scores above 0.96 on the HumanEval-CoT dataset and above 0.87 on the OpenEval-CoT dataset. In contrast, the majority of \ellLM s scored below 0.6 in both data sets.
In addition, in terms of the Consistency metric, gpt-3.5-turbo and gpt-4 achieved the best results on the OpenEval and HumanEval datasets, respectively, yet the cost of calling gpt-4 is 20 times higher than that of gpt-3.5-turbo. Therefore, we choose gpt-3.5-turbo as the Teacher Model in our study.

By further analysis based on varying models and parameter sizes, we can achieve interesting insights. Among CodeGen, StarCoder, CodeT5+, and CodeLlama models, StarCoder and CodeLlama showed potential in generating high-quality CoTs for guiding code generation. 
This could be attributed to the pre-training dataset of StarCoder and CodeLlama.
The pre-training dataset of StarCoder includes information from GitHub commits and issues while the pre-training dataset of CodeLlama includes information from contains 8\% of code-related natural language samples and 7\% of general-purpose natural language data. 
The additional sources of natural language information may contribute to a better code understanding, resulting in higher-quality CoTs.
In contrast, the encoder-decoder model CodeT5+ exhibited the lowest performance compared to the decoder-only models (i.e., CodeGen and StarCoder). This performance difference may be attributed to the model's architecture. In the few-shot scenario, the encoder-decoder model may perform worse than the decoder model on the CoT generation task.

Regarding parameter scale, only the StarCoder model exhibited a positive correlation between larger parameter scale and better performance. The performance of the CodeGen and CodeT5+ models did not strictly align with the scale of parameters. Only the 2B version of these models demonstrated improved performance.


\begin{tcolorbox}[width=1.0\linewidth, title={Summary of RQ1}]
The majority of \ellLM s face challenges in generating high-quality CoTs for guiding code generation independently.
\end{tcolorbox}

\begin{table*}[ht]
 \caption{Performance comparison of different base models to generate CoTs on the HumanEval-CoT and OpenEval-CoT datasets (Here COTTON is based on CodeLlama-7b).}
     \centering
\scalebox{0.92}{
\begin{tabular}{c|c|c|ccccccc}
  \toprule
\textbf{Corpus} & \textbf{Base Model} & \textbf{Model Size} & \textbf{BLEU-1} & \textbf{BLEU-2} & \textbf{BLEU-3} & \textbf{BLEU-4} & \textbf{Meteor} & \textbf{Rouge-L} & \textbf{Consistency}\\
  \midrule
\multirow{12}{*}{HumanEval-CoT}
& CodeBERT & 173M & 46.35 & 38.79 & 33.10 & 28.81 & 27.52 & 50.66 & 29.27\\
& GraphCodeBERT & 173M & 47.32 & 39.75 & 34.17 & 30.08 & 27.91 & 50.68 & 33.54\\
& CodeGPT & 124M & 26.91 & 47.96 & 41.14 & 36.13 & 31.91 & 52.95 & 57.32 \\
& CodeGPT-adapter & 124M & 54.10 & 45.19 & 38.42 & 33.40 & 30.56 & 51.04 & 52.44\\
& PLBART & 139M & 42.95 & 35.07 & 29.13 & 24.81 & 24.25 & 33.85 & 21.34\\
& CodeT5 & 223M & 61.03 & 53.16 & 46.85 & 42.00 & 34.89 & 58.93 & 79.88 \\
& NatGen & 223M & 62.76 & 54.84 & 48.50 & 43.59 & 35.92 & 59.91 & 82.32\\
& CodeGeeX2 & 6B & 62.57 & 54.18 & 47.54 & 42.33 & 35.77 & 59.72 & 92.68\\
& LLama2 & 7B & \textbf{66.56} & 58.00 & 51.05 & 45.62 & 37.65 & 61.39 & 89.63\\
& {\tool} & 7B & 65.97 & \textbf{58.21} & \textbf{51.89} & \textbf{46.87} & \textbf{38.22} & \textbf{63.38} & \textbf{93.29}\\
  \midrule
\multirow{12}{*}{OpenEval-CoT}
& CodeBERT & 173M & 34.19 & 27.18 & 21.98 & 18.14 & 23.02 & 41.20 & 8.99\\
& GraphCodeBERT & 173M & 37.87 & 30.09 & 24.42 & 20.27 & 23.73 & 42.23 & 11.24\\
& CodeGPT & 124M & 48.70 & 40.20 & 33.58 & 28.62 & 28.23 & 46.19 & 35.96 \\
& CodeGPT-adapter & 124M & 49.91 & 41.41 & 34.76 & 29.64 & 28.99 & 46.90 & 42.13 \\
& PLBART & 139M & 45.54 & 37.19 & 30.90 & 26.23 & 25.48 & 34.73 & 17.98 \\
& CodeT5 & 223M & 57.65 & 50.18 & 44.29 & 39.63 & 33.81 & 55.84 & 65.17 \\
& NatGen & 223M & 60.25 & 52.75 & 46.87 & 42.19 & 35.19 & 57.86 & 58.99 \\
& CodeGeeX2 & 6B & 64.86 & 56.83 & 50.54 & 45.62 & 37.05 & 61.31 & 79.21 \\
& LLama2 & 7B & 64.89 & 56.98 & 50.67 & 45.54 & 37.24 & 60.40 & 71.91 \\
& {\tool} & 7B & \textbf{67.04} & \textbf{59.56} & \textbf{53.60} & \textbf{48.80} & \textbf{38.80} & \textbf{62.92} & \textbf{83.71} \\
  \bottomrule
\end{tabular}
}
 \label{tab:RQ2}
\end{table*}


\subsection{RQ2: Can {\tool} generate higher-quality CoTs}  
\label{sect:rq2}


We treat the CoT generation as a text generation problem (i.e., convert a user's functional requirements into CoT). 
To provide a comprehensive evaluation, we compare nine commonly used baselines, including  CodeBERT~\cite{feng2020codebert}, GraphCodeBERT~\cite{guo2020graphcodebert}, CodeGPT~\cite{lu2021codexglue}, CodeGPT-adapter~\cite{lu2021codexglue}, PLBART~\cite{ahmad2021unified}, CodeT5~\cite{wang2021codet5}, NatGen~\cite{chakraborty2022natgen}, Llama2~\cite{touvron2023llama}, and CodeGeeX2~\cite{zheng2023codegeex}.
To ensure a fair comparison, we perform full parameter fine-tuning for models with less than 1B parameters. For models with more than 1B parameters, we employ LoRA~\cite{hu2021lora} for parameter-efficient fine-tuning. 

\begin{itemize}
    \item \textbf{CodeBERT}. CodeBERT~\cite{feng2020codebert} is a pre-trained encoder-only model that can handle multi-modal inputs, such as natural language descriptions of code or code comments. It combines these inputs with code snippets to generate more accurate and complete representations of code.
    
    \item \textbf{GraphCodeBERT}. GraphCodeBERT~\cite{guo2020graphcodebert} is an extension of CodeBERT specifically designed to capture more fine-grained information about code structures and dependencies. It enhances the capabilities of CodeBERT by incorporating graph-based representations.
    
    \item \textbf{CodeGPT}. CodeGPT~\cite{lu2021codexglue} is a pre-trained decoder-only model employing a 12-layer Transformer Decoder architecture. It is trained with the same structure as GPT-2.
    
    \item \textbf{CodeGPT-adapter.} CodeGPT-adapter~\cite{lu2021codexglue} is an extension of CodeGPT that utilizes domain-adaptive learning. It is initialized with a pre-trained GPT-2 model and then continues pretraining on the code dataset.
    
    \item \textbf{PLBART.} PLBART~\cite{ahmad2021unified} is a sequence-to-sequence model pre-trained on a large collection of Java and Python functions and natural language descriptions via denoising autoencoding.
    
    \item \textbf{CodeT5.} CodeT5~\cite{wang2021codet5} is a unified pre-trained encoder-decoder Transformer model that considers token type information in code and better leverages the code semantics conveyed from developer-assigned identifiers.
    
    \item \textbf{NatGen.} NatGen~\cite{chakraborty2022natgen} is an extension of CodeT5 that exploits the bimodal and dual-channel nature of code information to learn the naturalizing of source code.
    
    \item \textbf{CodeGeeX2.} CodeGeeX2~\cite{zheng2023codegeex} is a multilingual code generation model, which is based on the ChatGLM2 architecture. 
    
    \item \textbf{Llama2.} Llama2~\cite{touvron2023llama} is an extension of Llama. Building on Llama, Llama2 increases the size of the pre-trained corpus by 40\%, doubles the context length of the model, and employs a grouped query attention mechanism.
\end{itemize}

\subsubsection{Automatic Evaluation.}
We first evaluate different base models on the HumanEval-CoT and OpenEval-CoT datasets. We consider different evaluation metrics, including BLEU-1, BLEU-2, BLEU-3, BLEU-4, Meteor, Rouge-L, and Consistency. 
The results are shown in Table~\ref{tab:RQ2}. 

For the HumanEval-CoT dataset, {\tool} consistently outperformed the other base models across various evaluation metrics. This indicates that {\tool} generates CoTs with higher lexical similarity compared to the base models. 
For example, based on the \textbf{METEOR} metric, {\tool} achieved a score of 0.38, while the other base models scored between 0.27 and 0.37. 
In terms of semantic similarity, {\tool} also outperformed the other base models on the HumanEval-CoT dataset. With a score of 0.93, {\tool} demonstrated a higher level of semantic similarity to the actual code compared to the scores ranging from 0.29 to 0.92 achieved by the other base models. 
Importantly, {\tool} even outperforms larger LLMs (i.e., InternLM 123B, ChatGLM 130B, and Gemini), indicating its effectiveness in generating high-quality CoTs.
We can find similar trends on the OpenEval-CoT dataset in Table~\ref{tab:RQ2}, where {\tool} achieves better performance compared to the other base models.  

Furthermore, we observe a correlation between the parameter scale of the models and the performance of the generated CoTs. Specifically, larger models (such as CodeGeeX2, Llama2, and CodeCoT) showed significant performance improvement in CoT generation compared to the other models. The larger parameter scales of these models enabled them to capture more complex patterns and dependencies in the data, leading to better CoT generation ability.


\subsubsection{Human Evaluation.}

Although automatic evaluation metrics can offer valuable insights into the quality of generated CoTs, these metrics mainly concentrate on overlap or semantic similarity with the CoTs generated by the Teacher Model, possibly overlooking whether the generated CoT is indeed inspiring or offers meaningful guidance to developers.
Furthermore, the enhancements in Table~\ref{tab:RQ2} brought about by {\tool} over CodeGeeX2 and LLama2 may not be immediately apparent.

To further evaluate the quality of the generated CoTs, we conducted a human study using four groups of CoTs generated by different base models (CodeGeeX2 and LLama2) that performed similarly based on the results in Table~\ref{tab:RQ1} and Table~\ref{tab:RQ2}. 
By involving human evaluators, we aimed to obtain a more comprehensive assessment of the CoTs and gain a deeper understanding of their practical implications.

\begin{figure}[htbp]
	\centering
	\includegraphics[width=0.48\textwidth]{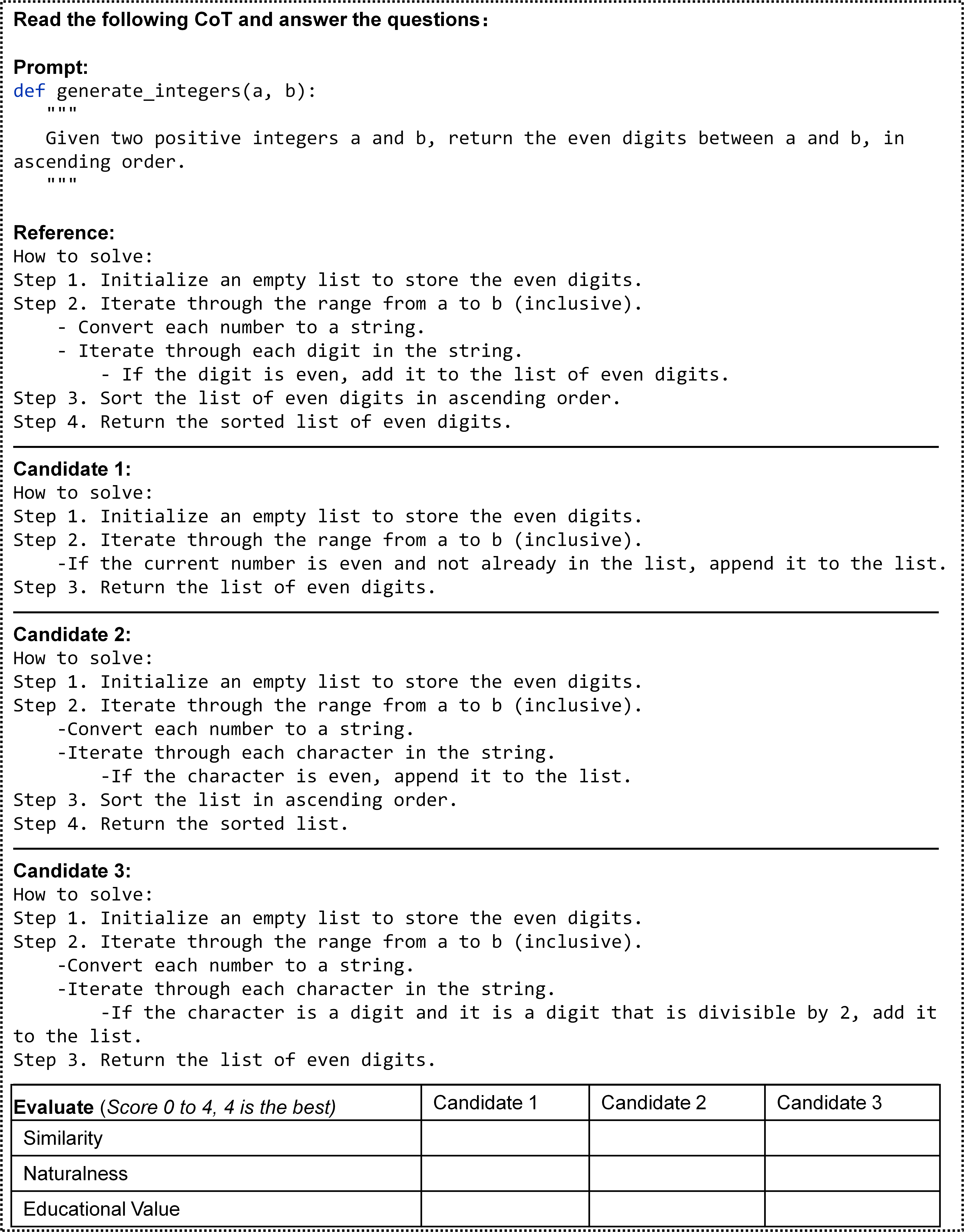}
	\caption{A sample questionnaire used in human study}
	\label{fig:humanExample}
\end{figure} 

In our human study, we adopted the methodology used in previous studies~\cite{wei2020retrieve, liu2023automated, li2023skcoder}. 
The quality of the generated CoTs was evaluated based on three aspects, as illustrated in \figurename~\ref{fig:humanExample}:

\begin{itemize}
    \item \textbf{Similarity}. This aspect measures the semantic similarity between the generated CoT and the reference.
    
    \item \textbf{Naturalness}. This aspect assessed the grammaticality and fluency of the generated CoT.
    
    \item \textbf{Educational Value}. This aspect evaluated whether volunteers could gain inspiration from the generated CoTs to solve problems, thereby measuring their educational value.
\end{itemize}

\begin{table}[htbp]
  \centering
  \caption{The average score and standard deviation (in parentheses) of human study.}
    \begin{tabular}{c|cccccc}
    \toprule
     & \textbf{Aspect}& \textbf{LLama2} & \textbf{CodeGeeX2} & \textbf{{\tool}} \\
    \midrule
    \multirow{3}{*}{HE}
    & Similarity & 2.847 (0.875) & 2.762 (0.918) & \textbf{3.038} (0.743) \\
    & Naturalness & 3.418 (0.842) & 3.375 (0.592) & \textbf{3.442} (0.713) \\
    & Educational Value & 2.985 (0.788) & 2.973 (0.812) & \textbf{3.102} (0.779) \\
    \midrule
    \multirow{3}{*}{OE} 
    & Similarity & 3.024 (0.645) & 2.976 (0.615) & \textbf{3.128} (0.734) \\
    & Naturalness & 3.521 (0.746) & 3.567 (0.789) & \textbf{3.691} (0.657) \\
    & Educational Value & 2.955 (0.858) & 3.002 (0.876) & \textbf{3.209} (0.891) \\
    \bottomrule
    \end{tabular}%
  \label{tab:humanResult}%
\end{table}%

\begin{table}[htbp]
  \centering
  \caption{The p-values between three aspects of the human study, and all the p-values are substantially smaller than 0.005.}
    \begin{tabular}{c|cccccc}
    \toprule
     & \textbf{Model}& \textbf{Similarity} & \textbf{Similarity} & \textbf{Educational Value} \\
    \midrule
    \multirow{2}{*}{HE}
    & Llama2 & 2.14e-5 & 1.85e-5 & 4.23e-5 \\
    & CodeGeeX2 & 5.68e-5 & 3.62e-5 & 1.45e-5 \\
    \midrule
    \multirow{2}{*}{OE} 
    & Llama2 & 1.53e-4 & 4.28e-4 & 8.85e-4 \\
    & CodeGeeX2 & 1.68e-4 & 4.29e-4 & 3.67e-4 \\
    \bottomrule
    \end{tabular}%
  \label{tab:humanResult-pvalue}%
\end{table}%

\begin{table*}[t]
\caption{The performance of code generation models with or without the CoTs generated by different self sot prompt methods on the HumanEval(HE), HumanEval-plus(HE-p), and OpenEval(OE) datasets}
     \centering
\begin{tabular}{c|c|c|ccccc}
  \toprule
\multirow{2}{*}{\textbf{Corpus}} & \multirow{2}{*}{\textbf{Model}} & \multirow{2}{*}{\textbf{Pass@1}} & 
\multicolumn{4}{c}{\textbf{CoT-Pass@1} (different CoT generation methods)} \\
&  &  & Self-CoT & Think Step by Step & Self-planning & SCoT & {\tool}\\
  \midrule
\multirow{10}{*}{HE} 
& CodeGen 350M & 14.63 & 12.80 & 15.85 ($\uparrow$ 8.34\%) & \underline{18.29} ($\uparrow$ 25.02\%) & 15.85 ($\uparrow$ 8.34\%) & \textbf{20.73} ($\uparrow$ 41.70\%) \\
& CodeGen 2B & 25.61 & 25.00 & \underline{27.44} ($\uparrow$ 7.15\%) & 26.22 ($\uparrow$ 2.38\%) & 26.83 ($\uparrow$ 4.76\%) & \textbf{34.76} ($\uparrow$ 35.73\%)\\
& CodeGen 6B & 27.44 & \underline{31.10} ($\uparrow$ 13.34\%) & 29.88 ($\uparrow$ 8.89\%) & 28.66 ($\uparrow$ 4.45\%) & 30.49 ($\uparrow$ 11.12\%) & \textbf{39.63} ($\uparrow$ 44.42\%) \\
& StarCoder 1B & 12.80 & 12.80 & 15.85 ($\uparrow$ 23.83\%) & \underline{17.68} ($\uparrow$ 38.13\%) & 14.63 ($\uparrow$ 14.30\%) & \textbf{25.00} ($\uparrow$ 95.31\%) \\
& StarCoder 3B & 17.07 & \underline{21.34} ($\uparrow$ 25.01\%) & 17.07 & 20.12 ($\uparrow$ 17.87\%) & 20.12 ($\uparrow$ 17.87\%) & \textbf{30.49} ($\uparrow$ 78.62\%) \\
& StarCoder 7B & 21.95 & \underline{29.88} ($\uparrow$ 36.13\%) & 28.05 ($\uparrow$ 27.79\%) & 25.61 ($\uparrow$ 16.67\%) & \underline{29.88} ($\uparrow$ 36.13\%) & \textbf{37.20} ($\uparrow$ 69.48\%) \\
& CodeT5+ 220M & 12.20 & \underline{14.02} ($\uparrow$ 14.92\%) & \underline{14.02} ($\uparrow$ 14.92\%) & 12.20 & 11.59 & \textbf{18.90} ($\uparrow$ 54.92\%) \\
& CodeT5+ 770M & 17.07 & 17.68 ($\uparrow$ 3.57\%) & 17.68 ($\uparrow$ 3.57\%) & 17.68 ($\uparrow$ 3.57\%) & \underline{18.90} ($\uparrow$ 10.72\%) & \textbf{26.83} ($\uparrow$ 57.18\%) \\
& CodeT5+ 2B & 23.78 & 25.00 ($\uparrow$ 5.13\%) & \underline{27.44} ($\uparrow$ 15.39\%) & 26.22 ($\uparrow$ 10.26\%) & 26.22 ($\uparrow$ 10.26\%) & \textbf{30.49} ($\uparrow$ 28.22\%) \\
& CodeT5+ 6B & 26.22 & 32.32 ($\uparrow$ 23.26\%) & 33.54 ($\uparrow$ 27.92\%) & \underline{34.15} ($\uparrow$ 30.24\%) & 32.93 ($\uparrow$ 25.59\%) & \textbf{42.68} ($\uparrow$ 62.78\%) \\
  \midrule
\multirow{10}{*}{OE} 
& CodeGen 350M & 7.30 & 8.99 ($\uparrow$ 23.15\%) & 8.43 ($\uparrow$ 15.48\%) & 7.87 ($\uparrow$ 7.81\%) & \underline{10.67} ($\uparrow$ 46.16\%) & \textbf{12.92} ($\uparrow$ 76.99\%) \\
& CodeGen 2B & 16.85 & 19.10 ($\uparrow$ 13.35\%) & 16.85 & \underline{20.22} ($\uparrow$ 20.00\%) & \underline{20.22} ($\uparrow$ 20.00\%) & \textbf{26.97} ($\uparrow$ 60.06\%) \\
& CodeGen 6B & 21.91 & 23.60 ($\uparrow$ 7.71\%) & 23.60 ($\uparrow$ 7.71\%) & 28.09 ($\uparrow$ 28.21\%) & \underline{29.78} ($\uparrow$ 35.92\%) & \textbf{33.71} ($\uparrow$ 53.86\%) \\
& StarCoder 1B & 8.99 & 10.67 ($\uparrow$ 18.69\%) & 8.99 & 11.24 ($\uparrow$ 25.03\%) & \underline{11.80} ($\uparrow$ 31.26\%) & \textbf{17.42} ($\uparrow$ 93.77\%) \\
& StarCoder 3B & 11.24 & \underline{16.85} ($\uparrow$ 49.91\%) & 14.61 ($\uparrow$ 29.98\%) & \textbf{19.10} ($\uparrow$ 69.93\%) & 14.61 ($\uparrow$ 29.98\%) & \textbf{19.10} ($\uparrow$ 69.93\%) \\
& StarCoder 7B & 23.03 & \underline{29.21} ($\uparrow$ 26.83\%) & 28.65 ($\uparrow$ 24.40\%) & 25.84 ($\uparrow$ 12.20\%) & 28.09 ($\uparrow$ 21.97\%) & \textbf{33.15} ($\uparrow$ 43.94\%) \\
& CodeT5+ 220M & 7.87 & 10.67 ($\uparrow$ 35.58\%) & \underline{12.36} ($\uparrow$ 57.05\%) & 10.67 ($\uparrow$ 35.58\%) & 8.43 ($\uparrow$ 7.12\%) & \textbf{13.48} ($\uparrow$ 71.28\%) \\
& CodeT5+ 770M & 9.55 & 11.24 ($\uparrow$ 17.70\%) & \underline{13.48} ($\uparrow$ 41.15\%) & \underline{13.48} ($\uparrow$ 41.15\%) & \underline{13.48} ($\uparrow$ 41.15\%) & \textbf{16.29} ($\uparrow$ 70.58\%) \\
& CodeT5+ 2B & 15.17 & \underline{19.66} ($\uparrow$ 29.60\%) & 16.85 ($\uparrow$ 11.07\%) & \underline{19.66} ($\uparrow$ 29.60\%) & 17.42 ($\uparrow$ 14.83\%) & \textbf{28.65} ($\uparrow$ 88.86\%) \\
& CodeT5+ 6B & 20.22 & 21.35 ($\uparrow$ 5.59\%) & 30.90 ($\uparrow$ 52.82\%) & 24.72 ($\uparrow$ 22.26\%) & \underline{31.46} ($\uparrow$ 55.59\%) & \textbf{35.39} ($\uparrow$ 75.02\%)\\
  \midrule
\multirow{10}{*}{HE-p} 
& CodeGen 350M & 15.24 & 12.80 & 16.46 ($\uparrow$ 8.01\%) & \underline{18.29} ($\uparrow$ 20.01\%) & 15.85 ($\uparrow$ 4.00\%) & \textbf{20.73} ($\uparrow$ 36.02\%) \\
& CodeGen 2B & 26.22 & 25.00 & \underline{27.44} ($\uparrow$ 4.65\%) & 26.22 & 26.83 ($\uparrow$ 2.33\%) & \textbf{35.37} ($\uparrow$ 34.90\%) \\
& CodeGen 6B & 27.44 & \underline{32.32} ($\uparrow$ 17.78\%) & 30.49 ($\uparrow$ 11.12\%) & 29.27 ($\uparrow$ 6.67\%) & 30.49 ($\uparrow$ 11.12\%) & \textbf{40.85} ($\uparrow$ 48.87\%) \\
& StarCoder 1B & 13.41 & 13.41 & 15.85 ($\uparrow$ 18.20\%) & \underline{17.68} ($\uparrow$ 31.84\%) & 14.63 ($\uparrow$ 9.10\%) & \textbf{26.22} ($\uparrow$ 95.53\%) \\
& StarCoder 3B & 17.07 & \underline{21.95} ($\uparrow$ 28.59\%) & 17.07 & 20.12 ($\uparrow$ 17.87\%) & 20.73 ($\uparrow$ 21.44\%) & \textbf{31.71} ($\uparrow$ 85.76\%) \\
& StarCoder 7B & 22.56 & 30.49 ($\uparrow$ 35.15\%) & 28.66 ($\uparrow$ 27.04\%) & 25.61 ($\uparrow$ 13.52\%) & \underline{31.10} ($\uparrow$ 37.85\%) & \textbf{38.41} ($\uparrow$ 70.26\%) \\
& CodeT5+ 220M & 12.20 & \underline{14.02} ($\uparrow$ 14.92\%) & \underline{14.02} ($\uparrow$ 14.92\%) & 12.20 & 11.59 & \textbf{19.51} ($\uparrow$ 59.92\%) \\
& CodeT5+ 770M & 17.68 & 17.68 & 17.68 & 17.68 & \underline{19.51} ($\uparrow$ 10.35\%) & \textbf{27.44} ($\uparrow$ 55.20\%) \\
& CodeT5+ 2B & 25.00 & 25.00 & \underline{28.05} ($\uparrow$ 12.20\%) & 27.44 ($\uparrow$ 9.76\%) & 26.22 ($\uparrow$ 4.88\%) & \textbf{31.71} ($\uparrow$ 26.84\%) \\
& CodeT5+ 6B & 26.83 & 32.93 ($\uparrow$ 22.74\%) & \underline{34.15} ($\uparrow$ 27.28\%) & 34.76 ($\uparrow$ 26.56\%) & 32.93 ($\uparrow$ 22.74\%) & \textbf{43.90} ($\uparrow$ 63.62\%) \\
  \bottomrule
\end{tabular}
 \label{tab:RQ3-1}
\end{table*}

For the human evaluation, we recruited six assessors (six postgraduate students) who were familiar with Python programming language. We selected all samples from the HumanEval-CoT dataset (164 samples) and the OpenEval-CoT dataset (178 samples) as the evaluation subjects. For each sample, we collected the ground truth CoT and three generated CoTs.

To ensure a thorough evaluation, we divided all the samples into three groups, with each group containing 114 samples. Each group was evaluated anonymously by two assessors in terms of similarity, naturalness, and educational value. The score for each aspect ranges from 0 to 4, with higher scores indicating higher quality and the final score is the average of two assessors’ scores.

A sample questionnaire is shown in \figurename~\ref{fig:humanExample}. To guarantee human study quality, the generated CoTs were presented in a random order, ensuring that the assessors had no knowledge of which approach generated the CoT. Moreover, the assessors were allowed to use the internet to look up any related concepts they were unfamiliar with.
Finally, we limited each assessor to evaluate only 20 samples in half a day. This was done to prevent fatigue and maintain a high level of concentration during the evaluation process. 

The results of the human study are summarized in Table~\ref{tab:humanResult}, which shows the average scores and standard deviations of all evaluated samples by the assessors. 
We can observe that the proposed approach {\tool} outperforms Llama2 and CodeGeeX2 in terms of similarity, naturalness, and educational value.
To determine the statistical significance of these differences, we conducted Fisher’s exact test~\cite{fisher1922interpretation} and observed a statistically significant disparity in Table~\ref{tab:humanResult-pvalue} (i.e., $p$-value $< 0.05$). 
Furthermore, we utilized Fleiss Kappa~\cite{fleiss1971measuring} to assess the agreement among the six assessors. The overall Kappa value based on the comparison results is 0.7, signifying substantial agreement among the assessors.

These findings provide further evidence of the effectiveness of {\tool} in generating high-quality CoTs for guiding code generation. The higher average score values obtained by {\tool} indicate that it is capable of generating CoTs that are more grammatically correct, fluent, and valuable in terms of providing guidance and inspiration to developers. 
These results highlight the competitiveness of our proposed approach in generating CoTs that are not only grammatically sound but also valuable from an educational perspective.

\begin{tcolorbox}[width=1.0\linewidth, title={Summary of RQ2}]
 {\tool} consistently outperforms the state-of-the-art base models 
 in  
generating CoTs 
in terms of lexical and semantic similarity, as well as its ability to provide valuable guidance and inspiration to developers.
\end{tcolorbox}


\begin{table*}[ht]
\caption{The performance of code generation models with or without the CoTs generated by different methods on the HumanEval(HE), HumanEval-plus(HE-p), and OpenEval(OE) datasets (Here COTTON is based on CodeLlama-7b).}
     \centering
\begin{tabular}{c|c|c|ccccc}
  \toprule
\multirow{2}{*}{\textbf{Corpus}} & \multirow{2}{*}{\textbf{Model}} & \multirow{2}{*}{\textbf{Pass@1}} & 
\multicolumn{4}{c}{\textbf{CoT-Pass@1} (different CoT generation methods)} \\
&  &  & Gemini & CodeLlama & ChatGLM & Teacher & {\tool}\\
  \midrule
\multirow{10}{*}{HE} 
& CodeGen 350M & 14.63 & \textbf{24.39} ($\uparrow$ 66.71\%) & 18.90 ($\uparrow$ 29.19\%) & 18.29 ($\uparrow$ 25.02\%) & \textbf{24.39} ($\uparrow$ 66.71\%) & \underline{20.73} ($\uparrow$ 41.70\%) \\
& CodeGen 2B & 25.61 & \underline{37.20} ($\uparrow$ 45.26\%) & 31.10 ($\uparrow$ 21.44\%) & 31.71 ($\uparrow$ 23.82\%) & \textbf{41.46} ($\uparrow$ 61.89\%) & 34.76 ($\uparrow$ 35.73\%)\\
& CodeGen 6B & 27.44 & \textbf{43.29} ($\uparrow$ 57.76\%) & 36.59 ($\uparrow$ 33.35\%) & 31.71 ($\uparrow$ 15.56\%) & \textbf{43.29} ($\uparrow$ 57.76\%) & \underline{39.63} ($\uparrow$ 44.42\%) \\
& StarCoder 1B & 12.80 & 21.95 ($\uparrow$ 71.48\%) & 17.68 ($\uparrow$ 38.13\%) & 17.68 ($\uparrow$ 38.13\%) & \textbf{28.66} ($\uparrow$ 123.91\%) & \underline{25.00} ($\uparrow$ 95.31\%) \\
& StarCoder 3B & 17.07 & \underline{34.15} ($\uparrow$ 100.06\%) & 25.61 ($\uparrow$ 50.03\%) & 25.61 ($\uparrow$ 50.03\%) & \textbf{39.63} ($\uparrow$ 132.16\%) & 30.49 ($\uparrow$ 78.62\%) \\
& StarCoder 7B & 21.95 & \underline{38.41} ($\uparrow$ 74.99\%) & 33.54 ($\uparrow$ 52.80\%) & 34.15 ($\uparrow$ 55.58\%) & \textbf{41.46} ($\uparrow$ 88.88\%) & 37.20 ($\uparrow$ 69.48\%) \\
& CodeT5+ 220M & 12.20 & \underline{19.51} ($\uparrow$ 59.92\%) & 18.29 ($\uparrow$ 49.92\%) & 16.46 ($\uparrow$ 34.92\%) & \textbf{23.17} ($\uparrow$ 89.92\%) & 18.90 ($\uparrow$ 54.92\%) \\
& CodeT5+ 770M & 17.07 & \underline{26.83} ($\uparrow$ 57.18\%) & 23.78 ($\uparrow$ 39.31\%) & 23.78 ($\uparrow$ 39.31\%) & \textbf{31.71} ($\uparrow$ 85.76\%) & \underline{26.83} ($\uparrow$ 57.18\%) \\
& CodeT5+ 2B & 23.78 & \textbf{38.41} ($\uparrow$ 61.52\%) & 28.05 ($\uparrow$ 17.96\%) & 29.27 ($\uparrow$ 23.09\%) & \textbf{38.41} ($\uparrow$ 61.52\%) & 30.49 ($\uparrow$ 28.22\%) \\
& CodeT5+ 6B & 26.22 & \underline{45.73} ($\uparrow$ 74.41\%) & 38.41 ($\uparrow$ 46.69\%) & 36.59 ($\uparrow$ 39.55\%) & \textbf{47.56} ($\uparrow$ 81.39\%) & 42.68 ($\uparrow$ 62.78\%) \\
  \midrule
\multirow{10}{*}{OE} 
& CodeGen 350M & 7.30 & \textbf{16.85} ($\uparrow$ 13.08\%) & 14.04 ($\uparrow$ 92.33\%) & 12.92 ($\uparrow$ 76.99\%) & \underline{15.17} ($\uparrow$ 107.81\%) & 12.92 ($\uparrow$ 76.99\%) \\
& CodeGen 2B & 16.85 & \textbf{30.34} ($\uparrow$ 80.06\%) & 21.91 ($\uparrow$ 30.03\%) & 25.28 ($\uparrow$ 50.03\%) & \underline{29.21} ($\uparrow$ 73.35\%) & 26.97 ($\uparrow$ 60.06\%) \\
& CodeGen 6B & 21.91 & \underline{34.27} ($\uparrow$ 56.41\%) & 29.21 ($\uparrow$ 33.32\%) & 32.02 ($\uparrow$ 46.14\%) & \textbf{37.64} ($\uparrow$ 71.79\%) & 33.71 ($\uparrow$ 53.86\%) \\
& StarCoder 1B & 8.99 & \underline{17.42} ($\uparrow$ 93.77\%) & 14.61 ($\uparrow$ 62.51\%) & 16.85 ($\uparrow$ 87.43\%) & \textbf{19.66} ($\uparrow$ 118.69\%) & \underline{17.42} ($\uparrow$ 93.77\%) \\
& StarCoder 3B & 11.24 & 18.54 ($\uparrow$ 64.95\%) & 15.17 ($\uparrow$ 34.96\%) & 14.61 ($\uparrow$ 29.98\%) & \underline{17.98} ($\uparrow$ 59.96\%) & \textbf{19.10} ($\uparrow$ 69.93\%) \\
& StarCoder 7B & 23.03 & \underline{34.83} ($\uparrow$ 51.24\%) & 28.65 ($\uparrow$ 24.40\%) & 29.78 ($\uparrow$ 29.31\%) & \textbf{38.20} ($\uparrow$ 65.87\%) & 33.15 ($\uparrow$ 43.94\%) \\
& CodeT5+ 220M & 7.87 & \textbf{15.17} ($\uparrow$ 92.76\%) & 10.67 ($\uparrow$ 35.58\%) & 12.92 ($\uparrow$ 64.17\%) & \underline{14.04} ($\uparrow$ 78.40\%) & 13.48 ($\uparrow$ 71.28\%) \\
& CodeT5+ 770M & 9.55 & \textbf{21.35} ($\uparrow$ 123.56\%) & 16.85 ($\uparrow$ 76.44\%) & 17.98 ($\uparrow$ 88.27\%) & \underline{20.22} ($\uparrow$ 111.73\%) & 16.29 ($\uparrow$ 70.58\%) \\
& CodeT5+ 2B & 15.17 & \underline{28.65} ($\uparrow$ 88.86\%) & 24.72 ($\uparrow$ 62.95\%) & 25.28 ($\uparrow$ 66.64\%) & \textbf{30.90} ($\uparrow$ 103.69\%) & \underline{28.65} ($\uparrow$ 88.86\%) \\
& CodeT5+ 6B & 20.22 & 33.15 ($\uparrow$ 63.95\%) & 28.09 ($\uparrow$ 38.92\%) & 32.02 ($\uparrow$ 58.36\%) & \textbf{36.52} ($\uparrow$ 80.61\%) & \underline{35.39} ($\uparrow$ 75.02\%)\\
  \midrule
\multirow{10}{*}{HE-p} 
& CodeGen 350M & 15.24 & \underline{24.39} ($\uparrow$ 60.04\%) & 19.51 ($\uparrow$ 28.02\%) & 18.29 ($\uparrow$ 20.01\%) & \textbf{25.00} ($\uparrow$ 64.04\%) & 20.73 ($\uparrow$ 36.02\%) \\
& CodeGen 2B & 26.22 & \underline{37.80} ($\uparrow$ 44.16\%) & 31.71 ($\uparrow$ 20.94\%) & 32.32 ($\uparrow$ 23.26\%) & \textbf{42.68} ($\uparrow$ 62.78\%) & 35.37 ($\uparrow$ 34.90\%) \\
& CodeGen 6B & 27.44 & \underline{43.90} ($\uparrow$ 59.99\%) & 37.20 ($\uparrow$ 35.57\%) & 32.32 ($\uparrow$ 17.78\%) & \textbf{44.51} ($\uparrow$ 62.21\%) & 40.85 ($\uparrow$ 48.87\%) \\
& StarCoder 1B & 13.41 & 22.56 ($\uparrow$ 68.23\%) & 18.29 ($\uparrow$ 36.39\%) & 18.29 ($\uparrow$ 36.39\%) & \textbf{29.27} ($\uparrow$ 118.27\%) & \underline{26.22} ($\uparrow$ 95.53\%) \\
& StarCoder 3B & 17.07 & \underline{34.76} ($\uparrow$ 103.63\%) & 25.61 ($\uparrow$ 50.03\%) & 26.22 ($\uparrow$ 53.60\%) & \textbf{40.85} ($\uparrow$ 139.31\%) & 31.71 ($\uparrow$ 85.76\%) \\
& StarCoder 7B & 22.56 & \underline{39.02} ($\uparrow$ 72.96\%) & 34.15 ($\uparrow$ 51.37\%) & 34.76 ($\uparrow$ 54.08\%) & \textbf{43.29} ($\uparrow$ 91.89\%) & 38.41 ($\uparrow$ 70.26\%) \\
& CodeT5+ 220M & 12.20 & \underline{20.12} ($\uparrow$ 64.92\%) & 18.90 ($\uparrow$ 54.92\%) & 17.07 ($\uparrow$ 39.92\%) & \textbf{23.78} ($\uparrow$ 94.92\%) & 19.51 ($\uparrow$ 59.92\%) \\
& CodeT5+ 770M & 17.68 & \underline{27.44} ($\uparrow$ 55.20\%) & 24.39 ($\uparrow$ 37.95\%) & 24.39 ($\uparrow$ 37.95\%) & \textbf{32.32} ($\uparrow$ 82.81\%) & \underline{27.44} ($\uparrow$ 55.20\%) \\
& CodeT5+ 2B & 25.00 & \underline{39.02} ($\uparrow$ 56.08\%) & 28.66 ($\uparrow$ 14.64\%) & 29.88 ($\uparrow$ 19.52\%) & \textbf{39.63} ($\uparrow$ 58.52\%) & 31.71 ($\uparrow$ 26.84\%) \\
& CodeT5+ 6B & 26.83 & \underline{46.34} ($\uparrow$ 72.72\%) & 39.02 ($\uparrow$ 45.43\%) & 37.20 ($\uparrow$ 38.65\%) & \textbf{48.78} ($\uparrow$ 81.81\%) & 43.90 ($\uparrow$ 63.62\%) \\
  \bottomrule
\end{tabular}
 \label{tab:RQ3-2}
\end{table*}


\begin{table*}[ht]
\caption{The performance of code generation models with or without the CoTs generated by {\tool} using different base models on the HumanEval(HE), HumanEval-plus(HE-p), and OpenEval(OE) datasets (Here COTTON is based on CodeLlama-7b).}
     \centering
\begin{tabular}{c|c|c|ccccc}
  \toprule
\multirow{2}{*}{\textbf{Corpus}} & \multirow{2}{*}{\textbf{Model}} & \multirow{2}{*}{\textbf{Pass@1}} & 
\multicolumn{4}{c}{\textbf{CoT-Pass@1} (different base models)} \\
&  &  & GraphCodeBERT & CodeGPT-adapter & NatGen & CodeGeeX2 & {\tool}\\
  \midrule
\multirow{10}{*}{HE} 
& CodeGen 350M & 14.63 & 15.24 ($\uparrow$ 4.17\%) & 15.85 ($\uparrow$ 8.34\%) & \underline{18.90} ($\uparrow$ 29.19\%) & \underline{18.90} ($\uparrow$ 29.19\%) & \textbf{20.73} ($\uparrow$ 41.70\%) \\
& CodeGen 2B & 25.61 & 25.61 & 24.39 & 32.32 ($\uparrow$ 26.20\%) & \underline{33.54} ($\uparrow$ 30.96\%) & \textbf{34.76} ($\uparrow$ 35.73\%)\\
& CodeGen 6B & 27.44 & 31.71 ($\uparrow$ 15.56\%) & 31.71 ($\uparrow$ 15.56\%) & 34.15 ($\uparrow$ 24.45\%) & \underline{35.96} ($\uparrow$ 19.51\%) & \textbf{39.63} ($\uparrow$ 44.42\%) \\
& StarCoder 1B & 12.80 & 17.07 ($\uparrow$ 33.36\%) & 15.24 ($\uparrow$ 19.06\%) & 15.85 ($\uparrow$ 23.83\%) & \underline{19.51} ($\uparrow$ 52.42\%) & \textbf{25.00} ($\uparrow$ 95.31\%) \\
& StarCoder 3B & 17.07 & 20.12 ($\uparrow$ 17.87\%) & 19.51 ($\uparrow$ 14.29\%) & 26.22 ($\uparrow$ 53.60\%) & \textbf{31.10} ($\uparrow$ 82.19\%) & \underline{30.49} ($\uparrow$ 78.62\%) \\
& StarCoder 7B & 21.95 & 28.05 ($\uparrow$ 27.79\%) & 24.39 ($\uparrow$ 11.12\%) & 30.49 ($\uparrow$ 38.91\%) & \underline{34.15} ($\uparrow$ 55.58\%) & \textbf{37.20} ($\uparrow$ 69.48\%) \\
& CodeT5+ 220M & 12.20 & 12.80 ($\uparrow$ 4.92\%) & 14.63 ($\uparrow$ 19.92\%) & 17.07 ($\uparrow$ 39.92\%) & \textbf{18.90} ($\uparrow$ 54.92\%) & \textbf{18.90} ($\uparrow$ 54.92\%) \\
& CodeT5+ 770M & 17.07 & 18.90 ($\uparrow$ 10.72\%) & 18.29 ($\uparrow$ 7.15\%) & 21.95 ($\uparrow$ 28.59\%) & \underline{25.00} ($\uparrow$ 46.46\%) & \textbf{26.83} ($\uparrow$ 57.18\%) \\
& CodeT5+ 2B & 23.78 & 27.44 ($\uparrow$ 15.39\%) & 25.00 ($\uparrow$ 5.13\%) & 28.05 ($\uparrow$ 17.96\%) & \textbf{30.49} ($\uparrow$ 28.22\%) & \textbf{30.49} ($\uparrow$ 28.22\%) \\
& CodeT5+ 6B & 26.22 & 31.71 ($\uparrow$ 20.94\%) & 32.32 ($\uparrow$ 23.26\%) & 37.80 ($\uparrow$ 44.16\%) & \underline{38.41} ($\uparrow$ 46.49\%) & \textbf{42.68} ($\uparrow$ 62.78\%) \\
  \midrule
\multirow{10}{*}{OE} 
& CodeGen 350M & 7.30 & 11.80 ($\uparrow$ 61.64\%) & 12.36 ($\uparrow$ 69.32\%) & 10.67 ($\uparrow$ 46.16\%) & \textbf{14.61} ($\uparrow$ 100.14\%) & \underline{12.92} ($\uparrow$ 76.99\%) \\
& CodeGen 2B & 16.85 & 17.42 ($\uparrow$ 3.38\%) & 22.47 ($\uparrow$ 33.35\%) & \underline{25.28} ($\uparrow$ 50.03\%) & 24.72 ($\uparrow$ 46.71\%) & \textbf{26.97} ($\uparrow$ 60.06\%) \\
& CodeGen 6B & 21.91 & 25.84 ($\uparrow$ 17.94\%) & 24.72 ($\uparrow$ 12.83\%) & 26.97 ($\uparrow$ 23.09\%) & \underline{28.09} ($\uparrow$ 28.21\%) & \textbf{33.71} ($\uparrow$ 53.86\%) \\
& StarCoder 1B & 8.99 & 11.24 ($\uparrow$ 25.03\%) & 10.67 ($\uparrow$ 18.69\%) & 11.80 ($\uparrow$ 31.26\%) & \underline{13.48} ($\uparrow$ 49.94\%) & \textbf{17.42} ($\uparrow$ 93.77\%) \\
& StarCoder 3B & 11.24 & 11.80 ($\uparrow$ 4.98\%) & 14.04 ($\uparrow$ 24.91\%) & 16.29 ($\uparrow$ 44.93\%) & \textbf{20.22} ($\uparrow$ 19.10\%) & \underline{19.10} ($\uparrow$ 69.93\%) \\
& StarCoder 7B & 23.03 & 22.47 & 25.28 ($\uparrow$ 9.77\%) & 25.28 ($\uparrow$ 9.77\%) & \underline{31.46} ($\uparrow$ 36.60\%) & \textbf{33.15} ($\uparrow$ 43.94\%) \\
& CodeT5+ 220M & 7.87 & 9.55 ($\uparrow$ 21.35\%) & 11.80 ($\uparrow$ 49.94\%) & \textbf{15.73} ($\uparrow$ 99.87\%) & \underline{15.17} ($\uparrow$ 92.76\%) & 13.48 ($\uparrow$ 71.28\%) \\
& CodeT5+ 770M & 9.55 & 15.17 ($\uparrow$ 58.85\%) & 13.48 ($\uparrow$ 41.15\%) & 13.48 ($\uparrow$ 41.15\%) & \textbf{16.85} ($\uparrow$ 76.44\%) & \underline{16.29} ($\uparrow$ 70.58\%) \\
& CodeT5+ 2B & 15.17& 17.42 ($\uparrow$ 14.83\%) & 20.79 ($\uparrow$ 37.05\%) & 23.03 ($\uparrow$ 51.81\%) & \underline{24.72} ($\uparrow$ 62.95\%) & \textbf{28.65} ($\uparrow$ 88.86\%) \\
& CodeT5+ 6B & 20.22 & 23.60 ($\uparrow$ 16.72\%) & 24.72 ($\uparrow$ 22.26\%) & 24.16 ($\uparrow$ 19.49\%) & \underline{29.78} ($\uparrow$ 47.28\%) & \textbf{35.39} ($\uparrow$ 75.02\%)\\
  \midrule
\multirow{10}{*}{HE-p} 
& CodeGen 350M & 15.24 & 15.24 & 15.85 ($\uparrow$ 4.00\%) & 18.90 ($\uparrow$ 24.02\%) & \underline{19.51} ($\uparrow$ 28.02\%) & \textbf{20.73} ($\uparrow$ 36.02\%) \\
& CodeGen 2B & 26.22 & 25.61 & 24.39 & 32.32 ($\uparrow$ 23.26\%) & \textbf{35.37} ($\uparrow$ 34.90\%) & \textbf{35.37} ($\uparrow$ 34.90\%) \\
& CodeGen 6B & 27.44 & 31.71 ($\uparrow$ 15.56\%) & 31.71 ($\uparrow$ 15.56\%) & 34.15 ($\uparrow$ 24.45\%) & \underline{36.59} ($\uparrow$ 33.35\%) & \textbf{40.85} ($\uparrow$ 48.87\%) \\
& StarCoder 1B & 13.41 & 17.07 ($\uparrow$ 27.29\%) & 15.24 ($\uparrow$ 13.65\%) & 15.85 ($\uparrow$ 18.20\%) & \underline{20.12} ($\uparrow$ 50.04\%) & \textbf{26.22} ($\uparrow$ 95.53\%) \\
& StarCoder 3B & 17.07 & 20.12 ($\uparrow$ 17.87\%) & 19.51 ($\uparrow$ 14.29\%) & 26.22($\uparrow$ 53.60\%) & \textbf{31.71} ($\uparrow$ 85.76\%) & \textbf{31.71} ($\uparrow$ 85.76\%) \\
& StarCoder 7B & 22.56 & 28.05 ($\uparrow$ 24.34\%) & 24.39 ($\uparrow$ 8.11\%) & 30.49 ($\uparrow$ 35.15\%) & \underline{34.76} ($\uparrow$ 54.08\%) & \textbf{38.41} ($\uparrow$ 70.26\%) \\
& CodeT5+ 220M & 12.20 & 12.80 ($\uparrow$ 4.92\%) & 14.63 ($\uparrow$ 19.92\%) & 17.07 ($\uparrow$ 39.92\%) & \textbf{19.51} ($\uparrow$ 59.92\%) & \textbf{19.51} ($\uparrow$ 59.92\%) \\
& CodeT5+ 770M & 17.68 & 18.90 ($\uparrow$ 6.90\%) & 18.29 ($\uparrow$ 3.45\%) & 21.95 ($\uparrow$ 24.15\%) & \underline{25.61} ($\uparrow$ 44.85\%) & \textbf{27.44} ($\uparrow$ 55.20\%) \\
& CodeT5+ 2B & 25.00 & 27.44 ($\uparrow$ 9.76\%) & 25.00 & 28.05 ($\uparrow$ 12.20\%) & \underline{31.10} ($\uparrow$ 24.40\%) & \textbf{31.71} ($\uparrow$ 26.84\%) \\
& CodeT5+ 6B & 26.83 & 31.71 ($\uparrow$ 18.19\%) & 32.32 ($\uparrow$ 20.46\%) & 37.80 ($\uparrow$ 40.89\%) & \underline{39.02} ($\uparrow$ 45.43\%) & \textbf{43.90} ($\uparrow$ 63.62\%) \\
  \bottomrule
\end{tabular}
 \label{tab:RQ3-3}
\end{table*}


\subsection{RQ3: Can \ellLM s  benefit from the generated CoT?} \label{sect:rq3}

\subsubsection{Effect of self-generated CoTs on \ellLM s.}

To assess the impact of generated CoTs on improving the code generation performance of LLMs (such as CodeGen, StarCoder, and CodeT5+ as discussed in Section~\ref{sect:cgm}), we initially compare different models in the Self-CoT scenario (i.e., the CoTs generated by the model itself using the few-shot method in \textbf{RQ1}).
Furthermore, since Self-planning~\cite{jiang2023self} and SCoT~\cite{li2023enabling} have demonstrated that well-designed self-CoT prompts can effectively improve the performance of large models in code generation, we also include Think step-by-step~\cite{kojima2022large}, Self-planning~\cite{jiang2023self}, and SCoT~\cite{li2023enabling} as baselines in our experiments.

We use the Pass@1 metric to evaluate the effectiveness of using CoTs in improving the performance of \ellLM s, which can help understand whether \ellLM s can effectively benefit from the instructions provided by CoT.
The results are presented in Table~\ref{tab:RQ3-1}, with the best result highlighted in boldface and the second-best result underscored.
We can observe that, under the guidance of the CoT generated by the Self-CoT method, the performance improvement of all \ellLM s is very limited, and even the performance of some \ellLM s may decrease. This finding supports the conclusion of \textbf{RQ1}, i.e.,  \ellLM s cannot independently generate high-quality CoTs.


\subsubsection{Effect of CoTs generated by different LMs on \ellLM s.}

We analyze the performance of \ellLM s when utilizing CoTs generated by CodeLlama, ChatGLM 130B, Gemini, and the Teacher model. Specifically, we compare the utility of CoTs generated by CodeLlama and the selected LLMs with those generated by {\tool} to determine whether {\tool}-generated CoTs are more beneficial for \ellLM s. The results are presented in Table~\ref{tab:RQ3-2}, with the best result highlighted in boldface and the second-best result underscored.

Our findings indicate that all \ellLM s can effectively leverage the guidance provided by CoTs to enhance the quality of generated code, assuming the CoTs' quality is ensured. This emphasizes the potential of utilizing CoTs to improve the performance of \ellLM s.

Furthermore, we examine the disparity between CoTs generated by {\tool} and those generated by the Teacher model. This comparison sheds light on how closely {\tool} can approximate the more advanced Teacher model in CoT generation. The experimental results reveal that {\tool} outperforms ChatGLM 130B and closely rivals Gemini and GPT-3.5-turbo in the CoT generation task for code generation.

\subsubsection{Effect of CoTs generated by different base models on \ellLM s.}

In our work, essentially we fine-tune a base model CodeLlama to obtain {\tool}. A legitimate question is why we choose CodeLlama, or whether other base models could yield better results.
To further investigate this question,  we take different base {\ellLM} models to tune them for CoT generation purposes. 
We then conduct experiments using the generated CoTs,  
where the results are presented in Table~\ref{tab:RQ3-3}, with the best result in boldface and the second-best result underscored.

Based on the results, it is observed that using CodeLlama as the base model leads to more significant performance improvements in code generation for the majority of \ellLM s, which justifies our choice and highlights the importance of choosing an appropriate base model when generating CoTs to enhance the code generation performance of \ellLM s.
By selecting the most effective base model, {\tool} can optimize the performance and effectiveness of \ellLM s in generating high-quality code.

\begin{tcolorbox}[width=1.0\linewidth, title={Summary of RQ3}]
\ellLM s can effectively leverage the guidance provided by CoTs. 
This emphasizes the potential of utilizing CoTs to enhance the performance of \ellLM s in code generation tasks.
\end{tcolorbox}
\section{Discussions}
\label{sec:discussion}

\subsection{Evaluating {\tool} on LLMs}
To further evaluate the capabilities of {\tool}, we conduct experiments to assess its impact on LLMs, specifically focusing on gpt-3.5-turbo. Our goal is to investigate whether {\tool} can enhance the performance of LLMs and to what extent this enhancement is achieved.

Gpt-3.5-turbo serves as a representative model in code generation and is one of the most state-of-the-art LLMs available. We consider its performance for code generation in zero-shot scenarios and then under the guidance of various COTs, including the CoTs generated by \tool, the CoTs self-generated by Self-planning~\cite{jiang2023self}, SCoT~\cite{li2023enabling}, and the Self-CoT method proposed in our study.

\begin{table}[ht]
 \caption{Evaluating {\tool} and other CoT prompt methods on gpt-3.5-turbo}
     \centering
\scalebox{0.95}{
\begin{tabular}{c|c|cccc}
  \toprule
\multirow{2}{*}{\textbf{Corpus}} & \multirow{2}{*}{\textbf{Pass@1}} & 
\multicolumn{4}{c}{\textbf{CoT-Pass@1}} \\
 &  &  COTTON & Self-Planning & SCoT & Self-CoT\\
  \midrule
HE & 56.10 & 74.39 & \textbf{77.44} & 76.83 & \textbf{77.44} \\
OE & 26.97 & 43.26 & 42.70 & 44.38 & \textbf{45.51} \\
HE-p & 57.32 & 76.22 & 78.66 & 78.66 & \textbf{79.27} \\
  \bottomrule
\end{tabular}
}
 \label{tab:gpt3_5}
\end{table}

Table \ref{tab:gpt3_5} shows the impact of {\tool} and other self-generated CoT methods on gpt-3.5-turbo. The experimental results show that the gpt-3.5-turbo model demonstrates a significant improvement in code generation performance when guided by various CoTs.
When comparing the zero-shot scenarios with the guided code generation, we observe a substantial increase in Pass@1, indicating that the incorporation of CoTs, including those generated by {\tool}, positively influences the model's ability to generate more accurate code. Notably, it even exceeds the performance of the GPT4 zero-shot scenario on the HumanEval dataset (67.0).
Moreover, the Self-CoT method proposed in our study can perform better than Self-Planning and SCoT on all datasets.


\begin{table*}[ht]
\caption{The performance comparison of zero-shot, fine-tuning and {\tool} with StarCoder series models.}
     \centering
\begin{tabular}{c|c|cccccc}
  \toprule
\multirow{2}{*}{} & \multirow{2}{*}{\textbf{StarCoder}} & \multicolumn{3}{c}{\textbf{Pass@1}} & \multicolumn{3}{c}{\textbf{CoT-Pass@1}}\\
 &  & Zero-shot & Fine-tune & Instruction-tune & {\tool}  & {\tool} w. Fine-tune & {\tool} w. Instruction-tune\\
  \midrule
\multirow{4}{*}{HE} 
& 1B & 12.80 & 14.63 ($\uparrow$ 14.30\%) & 16.46 ($\uparrow$ 28.59\%) & 25.00 ($\uparrow$ 95.31\%) & 25.00 ($\uparrow$ 95.31\%) & \textbf{28.05} ($\uparrow$ 119.14\%)\\
& 3B & 17.07 & 22.56 ($\uparrow$ 32.16\%) & 24.36 ($\uparrow$ 42.71\%) & 30.49 ($\uparrow$ 78.62\%) & 31.71 ($\uparrow$ 85.76\%) & \textbf{35.37} ($\uparrow$ 107.21\%) \\
& 7B & 21.95 & 25.61 ($\uparrow$ 16.67\%) & 26.83 ($\uparrow$ 22.23\%) & 37.20 ($\uparrow$ 69.48\%) & 38.41 ($\uparrow$ 74.99\%) & \textbf{40.24} ($\uparrow$ 83.33\%) \\
& 16B & 34.10 & 37.80 ($\uparrow$ 10.85\%) & 39.02 ($\uparrow$ 14.43\%) & 43.90 ($\uparrow$ 28.74\%) & 44.51 ($\uparrow$ 30.53\%) & \textbf{46.95} ($\uparrow$ 37.68\%)\\
  \midrule
\multirow{4}{*}{OE} 
& 1B & 8.99 & 8.43 & 11.24 ($\uparrow$ 25.03\%) & 17.42 ($\uparrow$ 93.77\%) & 17.98 ($\uparrow$ 100\%) & \textbf{20.22} ($\uparrow$ 124.92\%) \\
& 3B & 11.24 & 18.54 ($\uparrow$ 64.95\%) & 19.66 ($\uparrow$ 74.91\%) & 19.10 ($\uparrow$ 69.93\%) & 20.79 ($\uparrow$ 84.96\%) & \textbf{22.47} ($\uparrow$ 99.91\%)\\
& 7B & 23.03 & 24.16 ($\uparrow$ 4.91\%) & 26.40 ($\uparrow$ 14.63\%) & 33.15 ($\uparrow$ 43.94\%) & 33.71 ($\uparrow$ 46.37\%) & \textbf{35.96} ($\uparrow$ 56.14\%) \\
& 16B & 27.53 & 29.78 ($\uparrow$ 8.17\%) & 31.46 ($\uparrow$ 14.28\%) & 39.33 ($\uparrow$ 42.86\%) & 40.45 ($\uparrow$ 46.93\%) & \textbf{42.70} ($\uparrow$ 55.10\%)\\
  \midrule
\multirow{4}{*}{HE-p} 
& 1B & 13.41 & 14.63 ($\uparrow$ 9.10\%) & 16.46 ($\uparrow$ 22.74\%) & 26.22 ($\uparrow$ 95.53\%) & 26.83 ($\uparrow$ 100.07\%) & \textbf{28.66} ($\uparrow$ 113.72\%) \\
& 3B & 17.07 & 22.56 ($\uparrow$ 32.16\%) & 24.36 ($\uparrow$ 42.71\%) & 31.71 ($\uparrow$ 85.76\%) & 31.71 ($\uparrow$ 85.76\%) & \textbf{36.59} ($\uparrow$ 114.35\%) \\
& 7B & 22.56 & 26.22 ($\uparrow$ 16.22\%) & 26.83 ($\uparrow$ 22.23\%) & 38.41 ($\uparrow$ 70.26\%) & 38.41 ($\uparrow$ 74.99\%) & \textbf{40.24} ($\uparrow$ 83.33\%) \\
& 16B & 34.10 & 38.41 ($\uparrow$ 12.64\%) & 39.02 ($\uparrow$ 14.43\%) & 44.51 ($\uparrow$ 30.53\%) & 44.51 ($\uparrow$ 30.53\%) & \textbf{46.95} ($\uparrow$ 37.68\%)\\
  \bottomrule
\end{tabular}
 \label{tab:fine-tune}
\end{table*}

\subsection{Ablation study} 

\begin{figure}[ht]
	\centering
	\subfigure[Results in HumanEval-CoT]{%
		\includegraphics[width=0.24\textwidth]{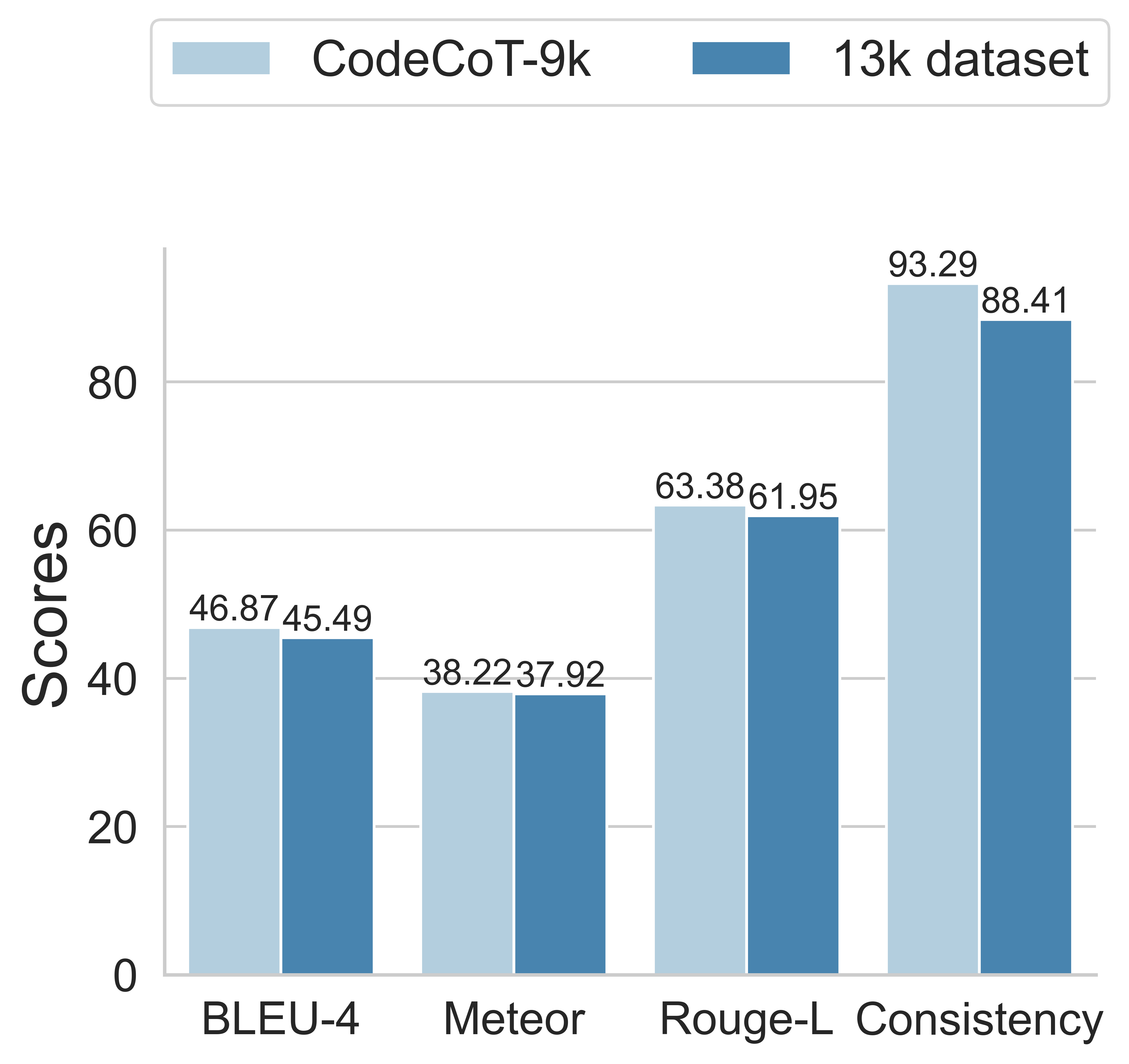}%
	}%
        \vspace{1mm}
	\subfigure[Results in OpenEval-CoT]{%
		\includegraphics[width=0.24\textwidth]{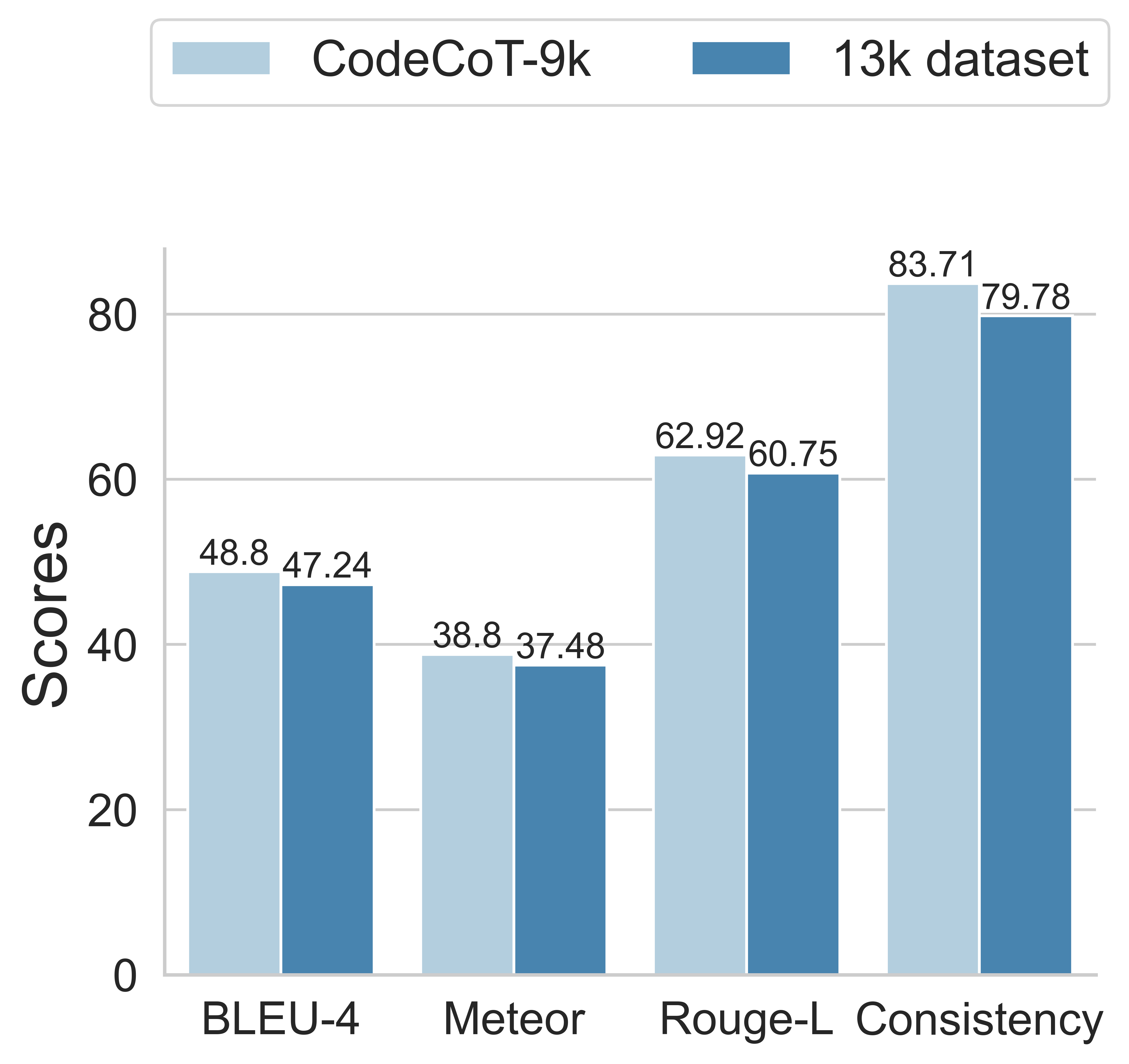}%
	}%
	\caption{The performance of {\tool} by whether using the Consistency Checker}
	\label{fig:ablation}
	\vspace{-1mm}
\end{figure}

The data collection methodology for {\dataset} relies on heuristic rules and multi-agent alignment. The heuristic rules (including Code Filtering, Document Filtering, and Similarity Filtering) have been widely used in training various LLMs (such as DeepSeek-Coder~\cite{guo2024deepseek}, StarCoder~\cite{li2023starcoder}, CodeT5+~\cite{wang2023codet5+}, and WizardCoder~\cite{luo2023wizardcoder}), and their importance has been demonstrated. 
The multi-agent alignment, particularly the Quality Checker, has been proven by Gunasekar et al.~\cite{gunasekar2023textbooks} to be beneficial for training models with educational value. However, the specific contribution of the Consistency Checker to the overall quality of the dataset remains unclear.
To evaluate the impact of the Consistency Checker, we conduct an ablation experiment. When the Consistency Checker is not applied, the dataset consists of 13k examples.
In this experiment, we train {\tool} separately on {\dataset} and a 13k dataset without Consistency Checker filtering.

The ablation results of the Consistency Checker are shown in \figurename~\ref{fig:ablation} where we can observe that the use of the Consistency Checker has a significant impact on the performance of {\tool}. Specifically, in terms of the Consistency metric, without the Consistency Checker for filtering, {\tool} shows a relative performance decrease of 5.23\% on HumanEval-CoT and 4.69\% on OpenEval-CoT.

\subsection{Comparing {\tool} with other tune methods}

In this subsection, we first compare {\tool} with traditional fine-tuning method using the StarCoder-series models as examples. 
The empirical study~\cite{zhuo2024astraios} shows that LoRA usually makes the most favorable trade-off between cost and performance in code generation tasks. Hence we mainly compare {\tool} with LoRA in our discussion.
\yg{During the traditional fine-tuning process, natural language is used as input, and code is generated as output.}

\yg{Given that existing state-of-the-art instruction-tuning methods usually involve additional data, such as human feedback~\cite{chen2023improving} and compiler feedback~\cite{dou2024stepcoder, yang2024intercode}, which can significantly improve the performance of downstream tasks of the model, we follow the method of Zheng et al.~\cite{zheng2024opencodeinterpreter} and add CoT in the form of multi-turn dialogues during instruction-tuning to explore the impact of instruction-tuning on the performance of the model.}

The results shown in Table~\ref{tab:fine-tune} demonstrate that the combination of StarCoder-7B with {\tool} can exceed the performance of StarCoder-16B in zero-shot scenarios and can even achieve results comparable to a fine-tuned/instruction-tuned StarCoder-16B model.
\yg{In addition, we find that multi-tune instruction-tuning method outperform traditional fine-tuning method, which is a finding similar to that of Zheng et al.~\cite{zheng2024opencodeinterpreter}.
We also find that {\tool}, as an independent CoT generation model, can be combined with the fine-tuned/instruction-tuned model to further improve the performance of the model in code generation tasks.}

Furthermore, when considering GPU across model deployment, inference, and training with float16 precision (as shown in Table~\ref{tab:gpu}), we carefully set parameters including a maximum input and output length of 256 and a batch size of 1.
It is important to highlight that fine-tuning a 7B model with LoRA already pushes the limits on a single consumer-grade GPU (GeForce RTX series). For individual developers, fine-tuning a 16B model would require multiple professional-grade GPUs (Quadro series and Tesla series), leading to significantly higher hardware costs compared to fine-tuning a 7B model.
\yg{In contrast, {\tool} enables performance enhancement across multiple models \emph{without} the need of fine-tuning individual model, which is highly desirable. }

\begin{table}[ht]
\caption{The comparison of GPU memory usage with StarCoder series models.}
     \centering
\begin{tabular}{c|ccc}
  \toprule
\textbf{StarCoder} & \textbf{Deploy} & \textbf{Inference} & \textbf{Training (Lora)}\\
  \midrule
1B & 2.34 GB & $\sim$3.50 GB & $\sim$12.00 GB \\
3B & 6.15 GB & $\sim$7.27 GB & $\sim$16.00 GB \\
7B & 14.72 GB & $\sim$15.90 GB & $\sim$23.00 GB \\
16B & 31.88 GB & $\sim$33.00 GB & $\sim$40.00 GB\\
  \bottomrule
\end{tabular}
 \label{tab:gpu}
\end{table}
\subsection{The impact of different decoding strategies} 

In the field of natural language processing~\cite{massarelli2020decoding, wiher2022decoding}, the performance of downstream tasks depends not only on the quality of the model itself but also on the decoding strategy in the prediction stage. 
In this subsection, we evaluate the impact of different decoding strategies on the performance of {\tool}. We consider four classical decoding strategies~\cite{hf2023generation} commonly used in text generation:
\begin{itemize}
\item \textbf{Greedy Search.} This strategy selects the token with the highest probability at each decoding step, resulting in a deterministic output.

\item\textbf{Multinomial Sampling.} This strategy samples tokens from the probability distribution at each decoding step, introducing randomness into the output.

\item\textbf{Beam Search.} This strategy maintains a beam of the top-$k$ partial sequences and selects the most probable complete sequence based on the joint probability.

\item\textbf{Constrastive Search.} This strategy~\cite{su2022contrastive} aims to optimize a trade-off between exploration and exploitation during decoding by considering both the model's predicted probability and the contrastive loss.
\end{itemize}

\begin{figure}[ht]
	\centering
	\subfigure[The experimental results in HumanEval-CoT]{%
		\includegraphics[width=0.48\textwidth]{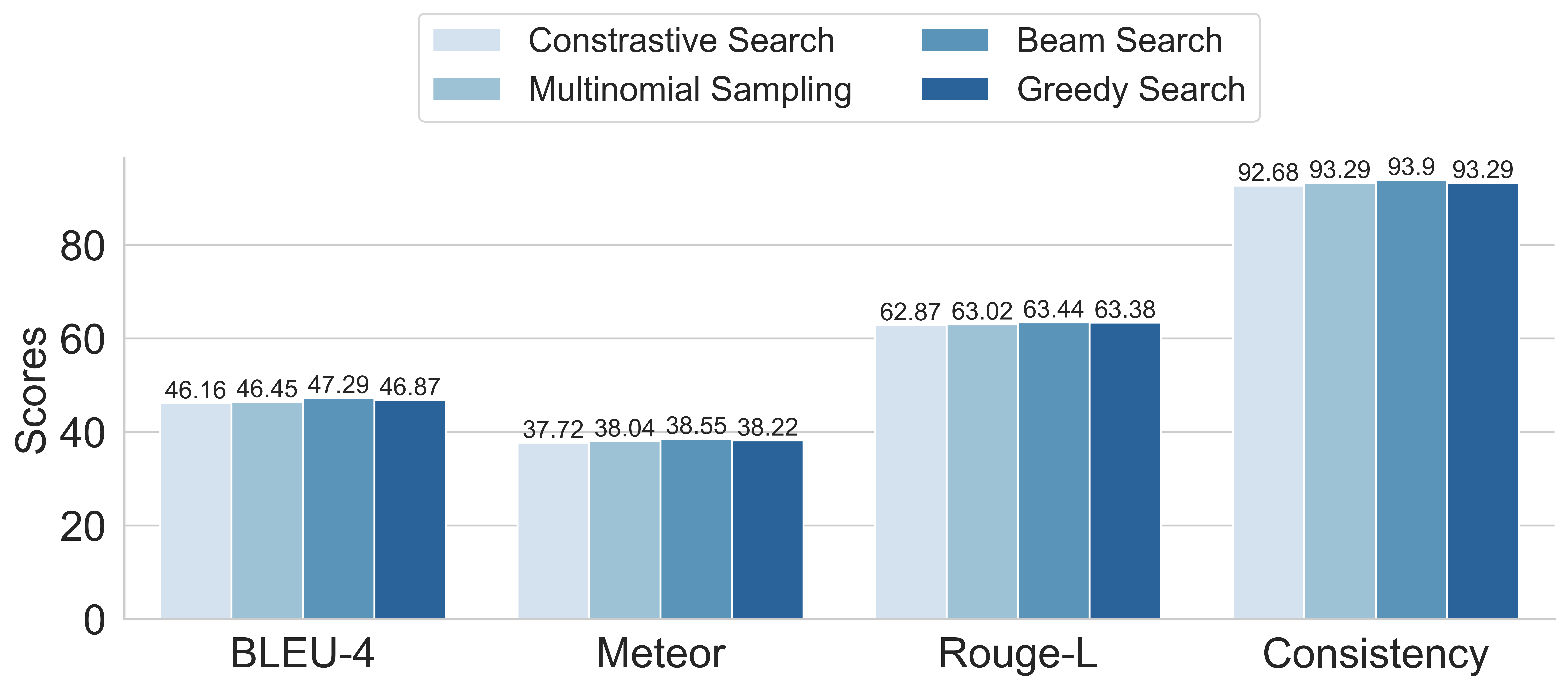}%
	}%
        \vspace{1mm}
	\subfigure[The experimental results in OpenEval-CoT]{%
		\includegraphics[width=0.48\textwidth]{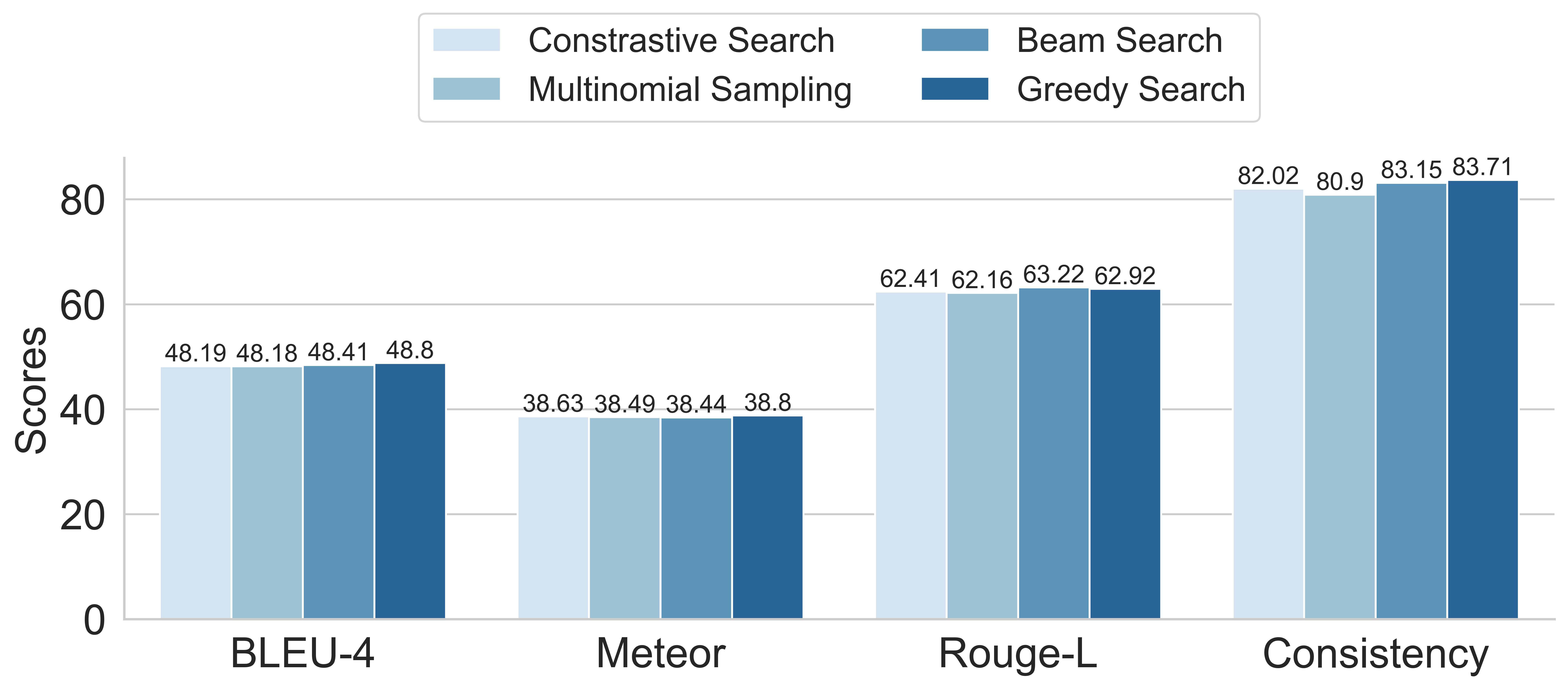}%
	}%
	\caption{The performance of {\tool} by using different decoding strategies}
	\label{fig:decoding}
	\vspace{-1mm}
\end{figure}

We show the evaluation results of using different decoding strategies in \figurename~\ref{fig:decoding} where we can find that the choice of decoding strategy does not have a significant impact on the performance of {\tool}.
Specifically, we only observe the minimal difference (i.e., variations of no more than 0.01) in terms of BLEU-4, METEOR, and Rouge-L performance metrics.
This suggests that {\tool} consistently performs well across different decoding strategies, generating high-quality COTs that are comparable in terms of these performance metrics. Therefore, the effectiveness of {\tool} in generating high-quality code does not heavily rely on the specific decoding strategy.

\subsection{The impact of different prompt methods} 

In this subsection, we will evaluate the impact of different prompt methods on the performance of {\tool}. We consider four classical prompt methods~\cite{peft} and one method without prompt:

\begin{itemize}
\item  \textbf{None:} This method refers to not using any prompt and letting the model generate CoTs without any specific guidance.

\item  \textbf{Prefix Tuning:} This method~\cite{li2021prefix} involves adding a task-specific sequence of vectors before the model input as a prefix.

\item  \textbf{Prompt Tuning:} This method~\cite{lester2021power} learns soft prompts to condition frozen language models for specific downstream tasks.

\item  \textbf{P-Tuning:} This method~\cite{liu2023gpt} utilizes trainable continuous prompt embeddings in combination with discrete prompts.

\item  \textbf{Alpaca Prompt.} This method~\cite{alpaca} designs a specific template for the prompts used to finetune LoRA models.
\end{itemize}

Table \ref{tab:prompt} shows the performance of {\tool} under different prompt methods on the HumanEval-CoT and OpenEval-CoT datasets, measured by BLEU-4 and Meteor scores. Among the prompt methods, Prefix Tuning consistently achieves the highest scores on both datasets. It demonstrates the effectiveness of adding task-specific prefixes to guide the model's responses. Prompt Tuning and Alpaca Prompt also show competitive performance, indicating the benefits of using learned soft prompts or specific prompt templates.

However, our proposed prompt template outperforms all the other prompt methods. It achieves the highest scores on both datasets, surpassing the baseline and other prompt-based approaches. This suggests that our method successfully leverages the strengths of the prompt methods while introducing additional improvements to enhance the model's performance.

\yg{This discussion shows that considering a variety of different prompt methods during model training can have unexpected benefits in improving model performance. Therefore, when targeting different downstream tasks, it is necessary to carefully try different prompt methods to maximize the potential of the model.}

\begin{table}[ht]
 \caption{The performance of {\tool} under different prompt methods}
     \centering
\begin{tabular}{c|cc|cc}
  \toprule
\multirow{2}{*}{\textbf{Method}} & \multicolumn{2}{c}{\textbf{HumanEval-CoT}} & 
\multicolumn{2}{c}{\textbf{OpenEval-CoT}} \\
& BLEU-4 & Meteor &  BLEU-4 & Meteor\\
  \midrule
None & 44.98 & 36.12 & 44.89 & 37.18 \\
Prefix Tuning & 46.25 & 37.48 & 47.72 & 38.23 \\
Prompt Tuning & 45.72 & 36.35 & 46.81 & 37.52 \\
P-Tuning & 45.43 & 36.87 & 46.23 & 37.67 \\
Alpaca Prompt & 45.56 & 37.05 & 47.36 & 37.90 \\
Ours & 46.87 & 38.22 & 48.80 & 38.80 \\
  \bottomrule
\end{tabular}
 \label{tab:prompt}
\end{table}

\subsection{The impact of different hyper-parameter setting} 

\begin{table}[ht]
 \caption{The performance of {\tool} under different hyper-parameter settings}
     \centering
\begin{tabular}{c|c|cc|cc}
  \toprule
\multirow{2}{*}{\textbf{$R$}} & \multirow{2}{*}{\textbf{$Alpha$}} & \multicolumn{2}{c}{\textbf{HumanEval-CoT}} & 
\multicolumn{2}{c}{\textbf{OpenEval-CoT}} \\
&  & BLEU-4 & Meteor &  BLEU-4 & Meteor\\
  \midrule
\multirow{3}{*}{4}
 & 8 & 45.74 & 37.33 & 46.75 & 37.68 \\
 & 16 & 46.29 & 37.87 & 47.20 & 38.12 \\
 & 32 & 46.83 & 38.08 & 48.10 & 38.23 \\
\midrule
\multirow{3}{*}{8}
 & 8 & 46.06 & 37.55 & 47.65 & 38.39 \\
 & 16 & 46.87 & \textbf{38.22} & \textbf{48.80} & \textbf{38.80} \\
 & 32 & \textbf{46.95} & 38.15 & 48.60 & 38.54 \\
\midrule
\multirow{3}{*}{16}
 & 8 & 45.52 & 37.12 & 46.88 & 37.25 \\
 & 16 & 46.41 & 37.91 & 47.95 & 37.41 \\
 & 32 & 46.68 & 38.02 & 48.45 & 38.12 \\
  \bottomrule
\end{tabular}
 \label{tab:hyperparam}
\end{table}

Recall that $R$ and $Alpha$ are the most important hyper-parameters for LoRA, where $R$ represents the LoRA attention dimension and $Alpha$ represents the scaling parameter for LoRA. 
In this subsection, we analyze the effect of different hyper-parameter settings on the performance of {\tool}.

The results are shown in 
Table~\ref{tab:hyperparam}, where we compare {\tool} with other different hyper-parameter settings, and the performance is evaluated in terms of the BLEU-4 and METEOR metrics for both HumanEval-CoT and OpenEval-CoT.
Based on the results, we find that 
a higher value of $Alpha$ tends to yield better performance.
On the other hand, the impact of $R$ on {\tool} performance is relatively less significant: varying its 
value does not lead to substantial performance improvement in terms of these evaluation metrics.



\subsection{Threats to Validity}

In this subsection, we analyze potential threats to the validity of our empirical study.

\noindent\textbf{Threats to Internal Validity.}
The first internal threat is the possibility of implementation faults in {\tool}. To mitigate this threat, we conduct a careful code inspection of the implementation and utilize well-established third-party libraries (such as PyTorch and Transformers).
The second internal threat is the implementation correctness of the considered baselines. To alleviate this threat, we implemented all baselines based on their shared models on platforms such as Hugging Face\footnote{\url{https://huggingface.co/models}}.

\noindent\textbf{Threats to External Validity.}
The main external threat lies in the datasets used in our study. To mitigate this threat, we start by selecting widely used open datasets as the raw data for CoT generation. We then apply three heuristic rule-based cleaning methods to preprocess these datasets. Moreover, we propose a multi-agent alignment-based method that leverages multiple agents to align and clean the data. 
For the code generation dataset, we select popular HumanEval and HumanEval-plus datasets. 
To ensure the generalization ability of {\tool} on another dataset, we also construct a new code generation dataset OpenEval, which provides a diverse and challenging set of programming tasks that could evaluate the model's ability to generate high-quality code.

\noindent\textbf{Threats to Construct Validity.}
The main construct threat is related to the metrics used in our automated evaluation. By treating CoT generation as a text generation problem, we utilize metrics based on term overlap (such as \textbf{BLEU}, \textbf{METEOR}, and \textbf{ROUGE-L}), which have been commonly used in similar studies on 
programming language processing~\cite{gao2020generating,liu2022sotitle}. 
Moreover, we introduce the \textbf{Consistency} metric to assess alignment based on the 
nature of the code generation task under investigation.
Furthermore, to ensure the generalizability of {\tool}, we collect a new code generation dataset OpenEval. We manually design five test cases for each problem to ensure as much quality as possible for OpenEval.
Finally, we employ \textbf{Pass@1} and \textbf{CoT-Pass@1} to evaluate the performance of code generation models. 
\yg{It is noticed that the Pass@k metric would significantly increase the computational cost of the evaluation, especially when k is large, in our study we decided to focus only on the Pass@1 metric in the evaluation, which reflects the actual ability of the model to generate the correct code for a given input in one go, which is the most critical aspect of functional correctness.}
To complement automated evaluation, we also conducted a human study to validate the effectiveness of our proposed approach further. To guarantee the quality of our human study, we follow the human study methodology used in previous studies of similar software engineering tasks~\cite{jiang2017automatically,wei2020retrieve}.

\section{Related Work}
\label{sec:related}
In this section, we summarize related studies on neural code generation and chain of thought generation. 

\subsection{Code Generation}

Earlier research on neural code generation predominantly relied on heuristic rules and expert systems, such as probabilistic grammar-based approaches~\cite{cohn2010inducing, allamanis2014mining} and domain-specific language techniques~\cite{gulwani2010dimensions, zan2023large}. 
However, these methods exhibited inflexibility and lacked scalability~\cite{zan2023large}. 
Other studies attempted to utilize static language models such as  n-gram~\cite{nguyen2013statistical, raychev2014code} and Hidden Markov models~\cite{sutskever2008recurrent}, but they struggled with sparse vector representations and failed to effectively capture long-term dependencies. 
Consequently, researchers turned their attention to neural networks, specifically CNN~\cite{liu2016automatic, sun2019grammar}, RNN~\cite{iyer2016summarizing, wan2018improving}, and LSTM~\cite{eriguchi2016tree, yin2017syntactic}, to model the relationship between natural language and code. 
In 2017, the Transformer model~\cite{vaswani2017attention}, initially designed for machine translation, was introduced and later applied to the task of neural code generation~\cite{mastropaolo2021studying, shah2021natural}. However, these deep learning models require a substantial amount of (labeled) natural language and code pairs for training and have inherent limitations in their capabilities.

With the development of language models, researchers have seen a diagram shift 
to pre-training and fine-tuning in neural code generation. At this stage, models with less than 1 billion parameters are commonly used. 
For instance, CodeBERT~\cite{feng2020codebert} has been utilized for automatic generation of exploit code~\cite{liguori2021evil, yang2023exploitgen}, while CodeGPT~\cite{lu2021codexglue} has been applied to automatic generation of Java and Python code.
PLBART~\cite{ahmad2021unified} and CodeT5~\cite{wang2021codet5} are pre-trained on multiple programming languages, making them suitable as base models for multi-language code generation tasks. 
Built upon the CodeT5 model, code of mixed programming styles (turducken) was generated where multi-task learning was used to enforce syntactic constraints~\cite{yang2023syntax}. 
In addition, models such as JuPyT5~\cite{chandel2022training} and PyMT5~\cite{chandel2022training} focus on the Python language specifically. They construct dedicated datasets and further enhance code generation performance through fine-tuning.

More recently, there has been a remarkable advancement in the development of LLMs with over 10 billion parameters. These models 
have demonstrated the ability to generate code in a zero-shot manner.
A notable milestone is Codex~\cite{chen2021evaluating}, which boasts an impressive 12 billion parameters. Codex has showcased its capabilities by solving 72.31\% of challenging Python programming problems created by humans. This model has also been successfully integrated into the commercial product Copilot.\footnote{\url{https://github.com/features/copilot}} 
Following the success of Codex, several other LLMs designed specifically for code generation tasks have emerged. 
For example, AlphaCode~\cite{li2022competition} focuses on solving competitive-level programming problems, while InCoder~\cite{fried2022incoder} supports code completion in arbitrary positions using bidirectional contexts. Additional models include CodeGen~\cite{nijkamp2022codegen, nijkamp2023codegen2}, 
StarCoder~\cite{li2023starcoder}, WizardCoder~\cite{luo2023wizardcoder}, OctoCoder~\cite{muennighoff2023octopack} and CodeLlama~\cite{roziere2023code},  
which have 
demonstrated their potential to solve complex programming problems and assist developers in various settings. 

However, fine-tuning these LLMs can be computationally expensive and resource-intensive. 
The focus of the current paper is the \ellLM s which we intend to use for code generation without updating parameters. 
In particular, we leverage the newly introduced CoT technology, which, as shown in the current paper, can be an effective means to improve the quality and accuracy of the generated code by \ellLM. 
Our study provides a cost-effective alternative to \ellLM s directly for code generation, making them more accessible for a wider range of applications and users.

\subsection{Chain of Thought Generation}

As the number of model parameters and volume of training data increase, LLMs have demonstrated impressive reasoning capabilities~\cite{wei2022emergent}. 
Recently, there has been a growing interest in enhancing the performance of LLMs in downstream tasks without the need to update model parameters.  One way to achieve this is to 
harness the inferential reasoning abilities of LLMs, 
a notable approach of which 
is the CoT prompting method
~\cite{wei2022chain}. This method enables LLMs to provide reliable answers through thoughtful consideration and explanation. 
Various approaches have been studied aiming to generate more accurate and reliable CoT possibly using LLMs themselves. For instance, He et al.~\cite{he2022rethinking} incorporate external knowledge as supporting information to generate more faithful CoT. 
Wang et al.~\cite{wang2022self} utilize self-consistency by generating multiple inference paths and answers, selecting the most frequently occurring answer as the final output, thereby improving the quality of CoT.
Creswell et al.~\cite{creswell2022selection} propose a selection-inference framework that employs LLMs as general processing modules. This framework alternates between selection and inference steps, generating a series of interpretable, causal reasoning steps leading to the final answer. 
Zhou et al.~\cite{zhou2022least} introduce the least-to-most prompting method, which breaks down complex problems into simpler subproblems and solves them sequentially.

The methods have limitations in relying on LLMs with more than 100 billion parameters. 
Researchers have developed smaller language models 
via knowledge distillation.
Ho et al.~\cite{ho2022large} introduced Fine-tune-CoT, which leverages GPT3(175B) as a reasoning teacher to enable complex reasoning in smaller models, thereby significantly reducing the model size requirements.
Li et al.~\cite{li2023symbolic} proposed Symbolic Chain-of-Thought Distillation (SCoTD), a method for training smaller student models using rationalizations sampled from a much larger teacher model. This approach distills the reasoning capabilities of the larger model into smaller models.
Shridhar et al.~\cite{shridhar2022distilling} 
utilized the step-by-step Chain-of-Thought (CoT) reasoning capabilities of larger models and distilled these abilities into smaller models. 

Our primary objective is to generate high-quality CoTs for code generation at a manageable cost. 
Apart from the domain-specific features  
in constructing CoT, 
we design a stand-alone model of relatively small sizes dedicated to CoT generation which can 
improve the performance 
code generation. 

\subsection{Multi-agent Collaboration}
Multi-agent collaboration refers to a framework where multiple autonomous agents interact with each other in a shared environment. 
These agents can be program scripts, software bots, or robots, each with their capabilities, goals, and perceptions~\cite{du2021survey}. 
They can communicate, cooperate, compete, or negotiate with each other to achieve complex goals or solve problems.
LLMs within multi-agent collaboration systems is an emerging area of research in the deep learning community~\cite{agashe2023evaluating}.

For instance, Zhang et al.~\cite{zhang2023proagent} proposed ProAgent, a system designed for robotic tasks that analyzes the current context, anticipates teammates' intentions, and formulates strategies based on this reasoning. 
Chen et al.~\cite{chen2022visualgpt} developed VisualGPT, which leverages vision-based Pretrained Language Models for image captioning tasks.
In terms of code generation, Huang et al.~\cite{huang2023agentcoder} proposed AgentCoder, a novel solution comprising a multi-agent framework with the programmer agent, the test designer agent, and the test executor agent.

Our proposed multi-agent alignment-based cleaning method contains Quality Checker, CoT Generator, and Consistency Checker.
Quality Checker serves the purpose of evaluating and filtering the educational significance of the code, while the CoT Generator focuses on generating the corresponding CoTs. Additionally, the Consistency Checker plays a crucial role in assessing and filtering the semantic consistency of the generated CoTs with the code. These three agents possess unique capabilities and task goals, enabling them to communicate and collaborate effectively to produce high-quality CoTs.
\section{Conclusion}
\label{sec:conclusion}

In this paper, we have introduced {\dataset} and {\tool}, which leverage lightweight language models (with parameters less than 10B) to generate high-quality CoT 
for code generation. 
We have demonstrated the effectiveness and efficiency of {\tool} in generating high-quality code CoTs. 
When equipped with these CoTs, 
existing \ellLM s have demonstrated significant performance improvements in code generation tasks. 
Our study enables \ellLM s to perform the code generation task better without additional updates to model parameters, 
making them more accessible to individual users. 

Potential future research includes further improving the performance of {\tool}. 
We plan to extend {\tool} to other programming languages and explore various potentially promising techniques, such as retrieval augmented generation, adversarial training and contrastive learning.
Moreover, based on the results of our study, it is important and promising to explore the lightweight language models for other software engineering tasks. 

\section*{Acknowledgements}
The authors would like to thank the editors and the anonymous reviewers for their insightful comments and suggestions, which can substantially improve the quality of this work.
This work was partially supported by the National Natural Science Foundation of China (NSFC, No.\ 62372232), the Fundamental Research Funds for the Central Universities (No.\ NG2023005), the Collaborative Innovation Center of Novel Software Technology and Industrialization, the Postgraduate Research \& Practice Innovation Program of Jiangsu Province (No.\ KYCX23\_0396), and the Short-term Visiting Program of Nanjing University of Aeronautics and Astronautics for Ph.D. Students Abroad (No.\ 240501DF16). T.\ Chen is partially supported by oversea grants from the State Key Laboratory of Novel Software Technology, Nanjing University (KFKT2022A03, KFKT2023A04).  

\bibliographystyle{IEEEtran}
\bibliography{main}

\begin{IEEEbiography}[{\includegraphics[width=1in,height=1.25in,clip]{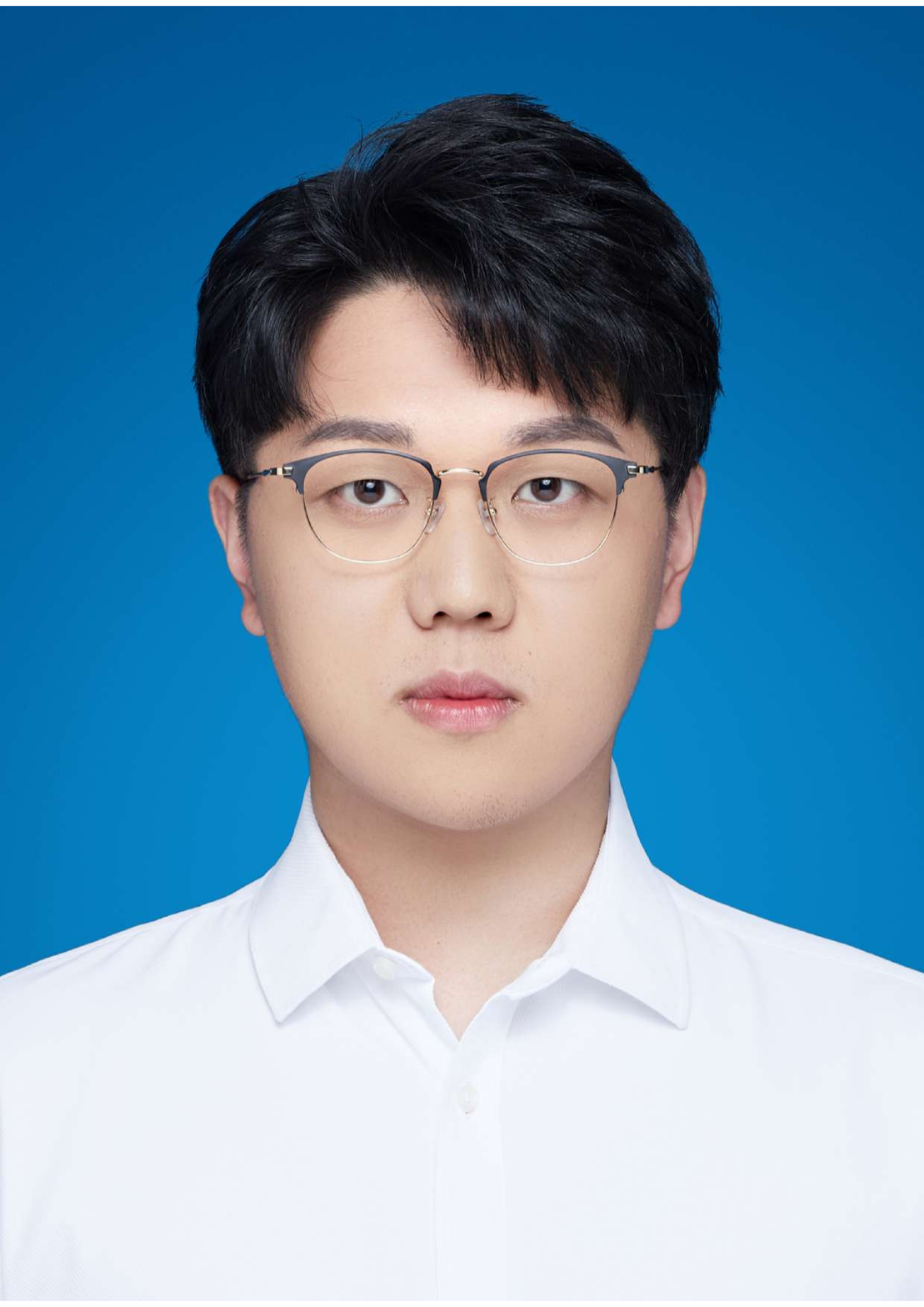}}]{Guang Yang}
received the M.D. degree in computer technology from Nantong University, Nantong, in 2022. Then he is currently pursuing the Ph.D degree at Nanjing University of Aeronautics and Astronautics, Nanjing.
His research interest is AI4SE and he has authored or co-authored more than 20 papers in refereed journals or conferences, such as ACM Transactions on Software Engineering and Methodology (TOSEM), Empirical Software Engineering, Journal of Systems and Software, International Conference on Software Maintenance and Evolution (ICSME), and International Conference on Software Analysis, Evolution and Reengineering (SANER).
More information about him can be found at: \url{https://ntdxyg.github.io/}
\end{IEEEbiography}

\begin{IEEEbiography}[{\includegraphics[width=1in,height=1.25in,clip]{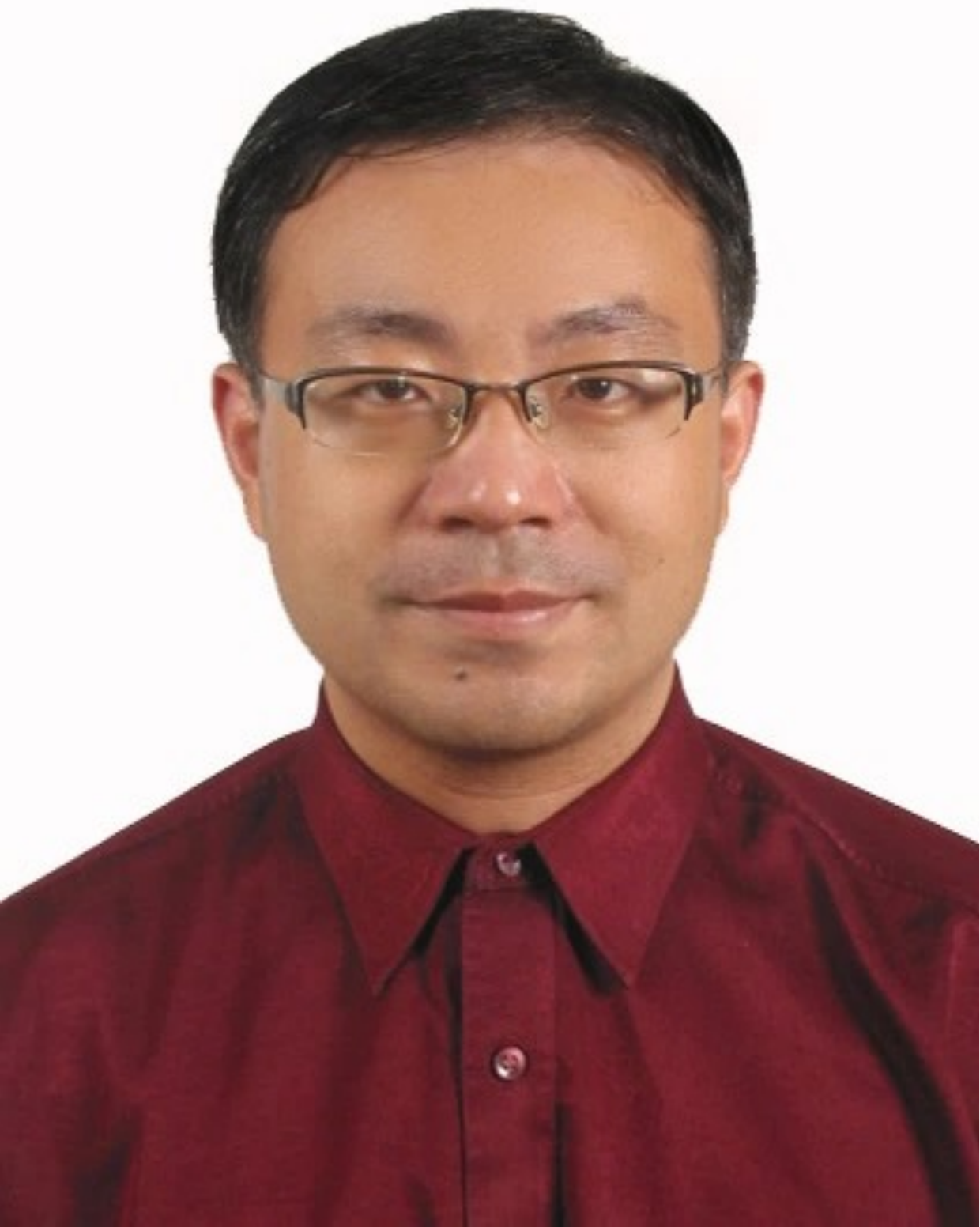}}]{Yu Zhou}
is a full professor in the College of Computer Science and Technology at Nanjing University of Aeronautics and Astronautics (NUAA). He received his BSc degree in 2004 and PhD degree in 2009, both in Computer Science from Nanjing University China. Before joining NUAA in 2011, he conducted PostDoc research on software engineering at Politechnico di Milano, Italy. From 2015-2016, he visited the SEAL lab at University of Zurich Switzerland, where he is also an adjunct researcher. His current research interests are mainly generative models for software engineering, software evolution analysis, mining software repositories, and reliability analysis. He has been supported by several national research programs in China. More information about him can be found at: \url{https://csyuzhou.github.io/}.
	
\end{IEEEbiography}

\begin{IEEEbiography}[{\includegraphics[width=1in,height=1.25in,clip]{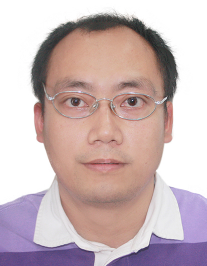}}]{Xiang Chen}
received the B.Sc. degree in the School of Management from Xi'an Jiaotong University, China in 2002. Then he received his M.Sc., and Ph.D. degrees in computer software and theory from Nanjing University, China in 2008 and 2011 respectively. He is currently an Associate Professor at the Department of Information Science and Technology, Nantong University, Nantong, China. He has authored or co-authored more than 120 papers in refereed journals or conferences, such as IEEE Transactions on Software Engineering, ACM Transactions on Software Engineering and Methodology, Empirical Software Engineering, Information and Software Technology, Journal of Systems and Software, International Conference on Software Engineering (ICSE), The ACM Joint European Software Engineering Conference and Symposium on the Foundations of Software Engineering (ESEC/FSE), International Conference Automated Software Engineering (ASE). His research interests include software engineering, in particular software testing and maintenance, software repository mining, and empirical software engineering. He received two ACM SIGSOFT distinguished paper awards in ICSE 2021 and ICPC 2023. He is the editorial board member of Information and Software Technology. More information about him can be found at: \url{https://xchencs.github.io/index.html}
\end{IEEEbiography}

\begin{IEEEbiography}[{\includegraphics[width=1in,height=1.25in,clip]{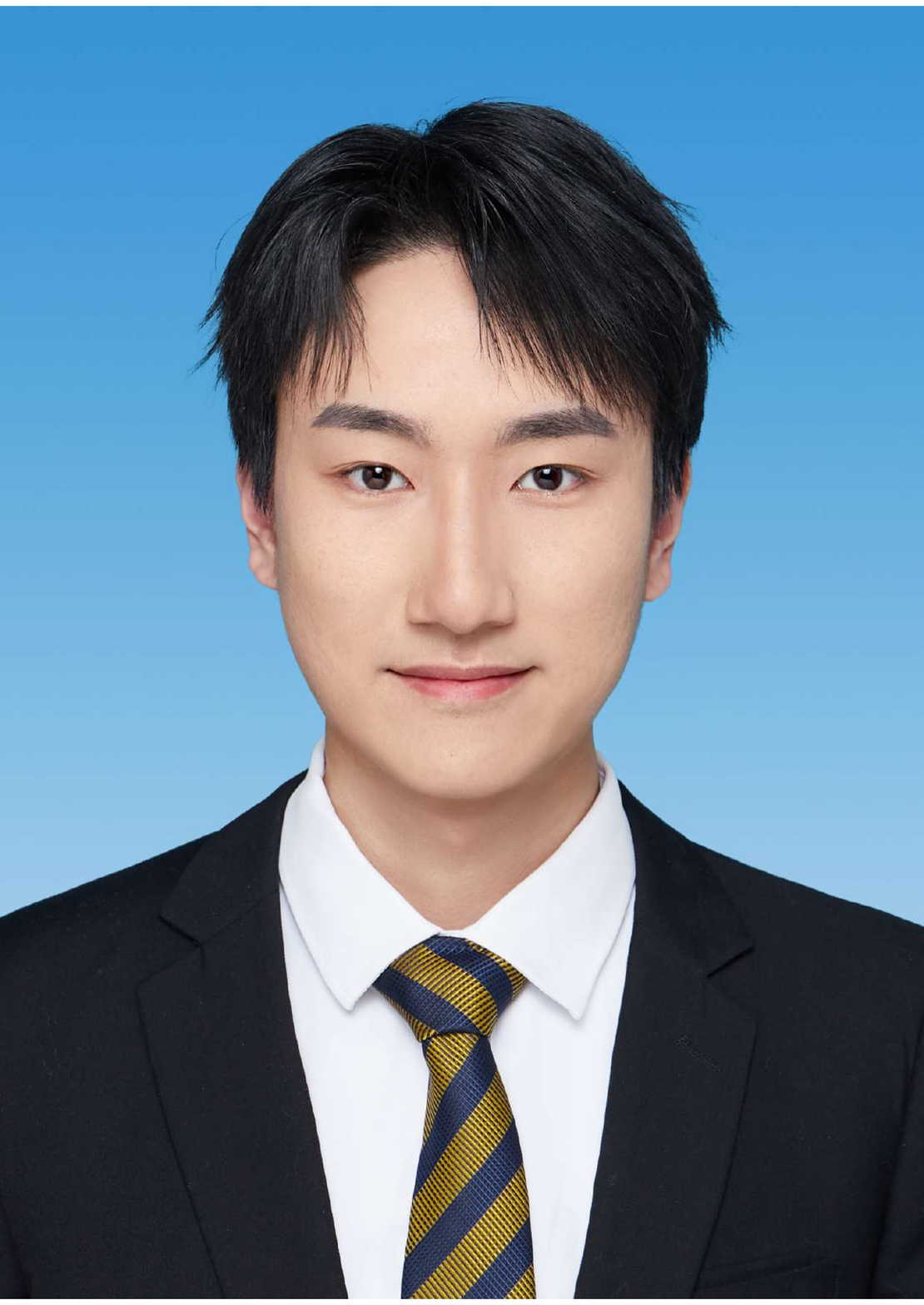}}]{Xiangyu Zhang}
	is currently pursuing a Master's degree at the College of Computer Science and Technology of Nanjing University of Aeronautics and Astronautics. His research interests include code generation and model interpretability.
\end{IEEEbiography}

\begin{IEEEbiography}[{\includegraphics[width=1in,height=1.25in,clip]{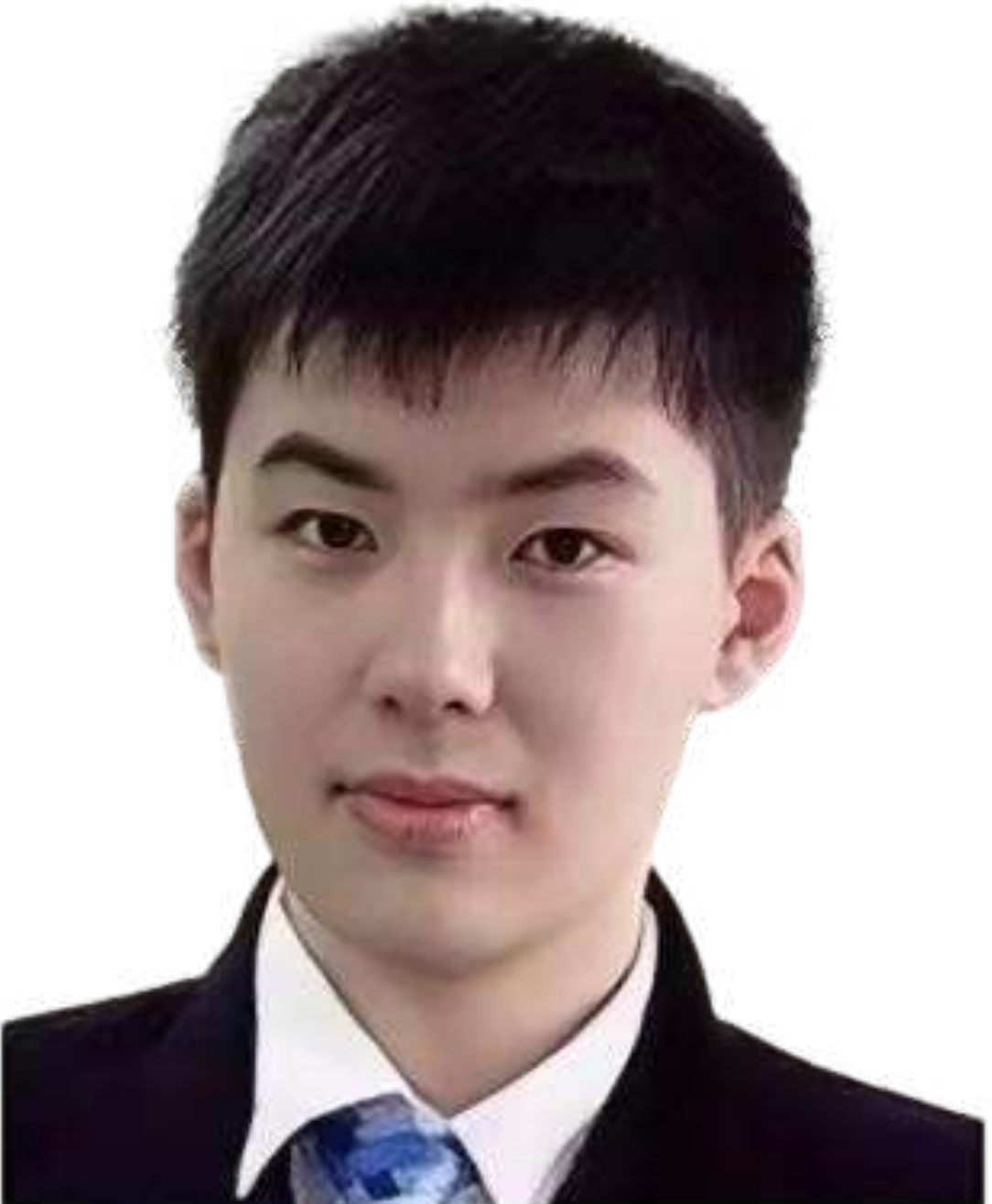}}]{Terry Yue Zhuo}
received the B.Sc. degree from the University of New South Wales, Sydney in 2021. Then he received the B.Sc. (Honours) degree from Monash University,  Australia, in 2022. He is currently pursuing a Ph.D. degree at Monash University and CSIRO's Data61, Australia. He is also working at CSIRO's Data61 as a research engineer and visiting the SOAR group at Singapore Management University. He has authored more than 15 papers in refereed journals or conferences, such as The Web Conference (WWW), ACM Transactions on Software Engineering and Methodology (TOSEM), Annual Meeting of the Association for Computational Linguistics (ACL) and Conference on Empirical Methods in Natural Language Processing (EMNLP). He received the best paper award in the Deep Learning for Code (DL4C) workshop in ICRL 2023. His research interests include empirical software engineering, code intelligence and responsible AI. More information about him can be found at \url{https://terryyz.github.io/}.
\end{IEEEbiography}

\begin{IEEEbiography}[{\includegraphics[width=1in,clip,keepaspectratio]{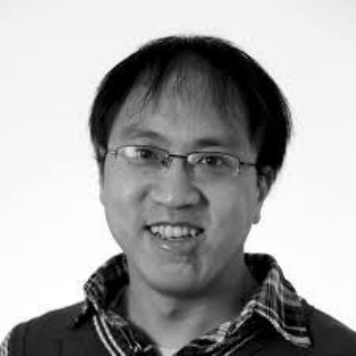}}]{Taolue Chen} received the Bachelor and Master degrees from Nanjing University, China, both in Computer Science. He was a junior researcher (OiO) at the Centrum Wiskunde \& Informatica (CWI) and acquired the PhD degree from the Vrije Universiteit Amsterdam, The Netherlands. He is currently a lecturer at the School of Computing and Mathematical Sciences, Birkbeck, University of London. He had been a postdoctoral researcher at University of Oxford (UK)  and University of Twente (NL). His research spans Software Engineering,  Program Language, Verification and Machine Learning. His present research focus is at the interface of software engineering and machine learning. He applies verification and programming language techniques to improve the trustworthiness of machine learning models. Meanwhile, he applies data-driven approaches to support software development. He has published over 140 papers in journals and conferences such as POPL, LICS, CAV, OOPSLA, ICSE, ESEC/FSE, ASE, ETAPS (TACAS, FoSSaCS, ESOP, FASE), NeurIPS, ICLR, IJCAI, AAAI, EMNLP and IEEE Transactions on Software Engineering (TSE), ACM Transactions on Software Engineering and Methodology (TOSEM), Empirical Software Engineering, ACM Transactions on Computational Logic (TOCL), Information and Computation, Logical Methods in Computer Science. He won the Best Paper Award of SETTA’20, the 1st Prize in the CCF Software Prototype Competition 2022, and the QF\_Strings (Single Query Track) at the International Satisfiability Modulo Theories  Competition 2023. He has served editorial board or program committee for various international journals and conferences. More information about him can be found at \url{https://chentaolue.github.io/}.
\end{IEEEbiography}

\end{document}